\documentclass[12pt]{article} 

\usepackage[english]{babel}
\usepackage{blindtext}
\usepackage{mathptmx}
\usepackage[scaled=.90]{helvet}
\usepackage{xcolor}
\usepackage{titlesec}
\titleformat{\chapter}[display]
  {\normalfont\sffamily\huge\bfseries\color{red}}
  {\chaptertitlename\ \thechapter}{1em}{\Huge}
  
\titleformat{\section}
  {\normalfont\sffamily\centering\fontsize{19pt}{12pt}\selectfont\bfseries\color{black}}
  {\thesection}{1em}{}

\titleformat{\subsection}
  {\normalfont\sffamily\centering\fontsize{20pt}{12pt}\selectfont\bfseries\color{black}}
  {\thesection}{1em}{}

\titleformat{\subsubsection}
  {\normalfont\sffamily\fontsize{15pt}{0pt}\selectfont\bfseries\color{black}}
  {\thesection}{10pt}{}

\titlespacing\section{0pt}{3pt plus 0pt minus 0pt}{9pt plus 0pt minus 0pt} 
\titlespacing\subsection{0pt}{-9pt plus 0pt minus 0pt}{1pt plus 0pt minus 0pt}
\titlespacing\subsubsection{0pt}{12pt plus 4pt minus 2pt}{3pt plus 2pt minus 2pt}

\usepackage[top=30mm, bottom=30mm, left=25mm, right=25mm]{geometry}

\usepackage{setspace}

\usepackage{latexsym}

\usepackage{amsmath}

\usepackage{amssymb}

\usepackage[font=small,labelfont=bf]{caption}

\usepackage{caption}

\usepackage{geometry} 

\usepackage{graphicx} 

\usepackage{float} 

\usepackage{wrapfig} 

\usepackage{lipsum} 

\usepackage{titlesec} 

\usepackage{fancyhdr}
\usepackage{textcomp} 

\usepackage[justification=justified,singlelinecheck=off]{caption}

\usepackage{graphicx}
\usepackage{caption}
\usepackage{subcaption}

\usepackage{threeparttable}
\usepackage{amsmath}
\usepackage{dcolumn}
\usepackage{multirow}
\usepackage{booktabs}
\newcolumntype{d}[1]{D{.}{.}{#1}}
\usepackage{longtable}

\begin{document}

\fancyfoot[C]{\thepage}

\renewcommand{\headrulewidth}{0pt}


\section*{A hybrid type Ia supernova with an early flash triggered by helium-shell detonation}

\subsection*{} 

\begin{spacing}{2.0}

\noindent Ji-an Jiang$^{1,2}$, Mamoru Doi$^{1,3,4}$, Keiichi Maeda$^{5,3}$, Toshikazu Shigeyama$^{4}$, Ken'ichi Nomoto$^{3}$, Naoki Yasuda$^{3}$, Saurabh~W. Jha$^{6}$, Masaomi Tanaka$^{7,3}$, Tomoki Morokuma$^{1,3}$, Nozomu Tominaga$^{8,3}$, \v{Z}eljko Ivezi\'c$^{9}$, Pilar Ruiz-Lapuente$^{10,11}$, Maximilian~D. Stritzinger$^{12}$, Paolo~A. Mazzali$^{13,14}$, Christopher Ashall$^{13}$, Jeremy Mould$^{15}$, Dietrich Baade$^{16}$, Nao Suzuki$^{3}$, Andrew~J. Connolly$^{9}$, Ferdinando Patat$^{16}$, Lifan Wang$^{17,18}$, Peter Yoachim$^{9}$, David Jones$^{19,20}$, Hisanori Furusawa$^{7}$, Satoshi Miyazaki$^{7,21}$

\noindent $^1$\emph{Institute of Astronomy, Graduate School of Science, The University of Tokyo, 2-21-1 Osawa, Mitaka, Tokyo 181-0015, Japan}

\noindent $^2$\emph{Department of Astronomy, Graduate School of Science, The University of Tokyo, 7-3-1 Hongo, Bunkyo-ku, Tokyo 113-0033, Japan} 

\noindent $^3$\emph{Kavli Institute for the Physics and Mathematics of the Universe (WPI), The University of Tokyo, 5-1-5 Kashiwanoha, Kashiwa, Chiba 277-8583, Japan}

\noindent $^4$\emph{Research Center for the Early Universe, Graduate School of Science, The University of Tokyo, 7-3-1 Hongo, Bunkyo-ku, Tokyo 113-0033, Japan}

\noindent $^5$\emph{Department of Astronomy, Kyoto University, Kitashirakawa-Oiwake-cho, Sakyo-ku, Kyoto 606-8502, Japan}

\noindent $^6$\emph{Department of Physics and Astronomy, Rutgers, The State University of New Jersey, 136 Frelinghuysen Road, Piscataway, NJ 08854, USA}

\noindent $^7$\emph{National Astronomical Observatory of Japan, 2-21-1 Osawa, Mitaka, Tokyo 181-8588, Japan}

\noindent $^8$\emph{Department of Physics, Faculty of Science and Engineering, Konan University, 8-9-1 Okamoto, Kobe, Hyogo 658-8501, Japan}

\noindent $^9$\emph{Department of Astronomy, University of Washington, Box 351580, Seattle, WA 98195-1580, USA}

\noindent $^{10}$\emph{Instituto de F\'isica Fundamental, Consejo Superior de Investigaciones Cient\'ificas, c/. Serrano 121, E-28006, Madrid, Spain}

\noindent $^{11}$\emph{Institut de Ci\`encies del Cosmos (UB-IEEC), c/. Mart\'i i Franqu\'es 1, E-08028, v, Spain}

\noindent $^{12}$\emph{Department of Physics and Astronomy, Aarhus University, Ny Munkegade 120, 8000 Aarhus C, Denmark}

\noindent $^{13}$\emph{Astrophysics Research Institute, Liverpool John Moores University, IC2, Liverpool Science Park, 146 Brownlow Hill, Liverpool L3 5RF, UK}

\noindent $^{14}$\emph{Max-Planck-Institut f\"ur Astrophysik, Karl-Schwarzschild-Str. 1, D-85748 Garching, Germany}

\noindent $^{15}$\emph{Centre for Astrophysics and Supercomputing, Swinburne University of Technology, Hawthorn, Vic 3122, Australia}

\noindent $^{16}$\emph{European Organisation for Astronomical Research in the Southern Hemisphere (ESO), Karl-Schwarzschild-Str. 2, 85748 Garching b. M\"unchen, Germany}

\noindent $^{17}$\emph{George P. and Cynthia Woods Mitchell Institute for Fundamental Physics and Astronomy, Department of Physics and Astronomy, Texas A\&M University, 4242 TAMU, College Station, TX 77843, USA}

\noindent $^{18}$\emph{Purple Mountain Observatory, Chinese Academy of Sciences, Nanjing 210008, China}

\noindent $^{19}$\emph{Instituto de Astrof\'isica de Canarias, E-38205 La Laguna, Tenerife, Spain}

\noindent $^{20}$\emph{Departamento de Astrof\'isica, Universidad de La Laguna, E-38206 La Laguna, Tenerife, Spain}

\noindent $^{21}$\emph{SOKENDAI (The Graduate University for Advanced Studies), Mitaka, Tokyo, 181-8588, Japan}

\end{spacing}

\subsection*{} 

\begin{spacing}{2.0}

\textbf{Type Ia supernovae (SNe Ia) arise from the thermonuclear explosion of carbon-oxygen white dwarfs$^{1,2}$. Though the uniformity of their light curves makes them powerful cosmological distance indicators$^{3,4}$, long-standing issues remain regarding their progenitors and explosion mechanisms$^{2,5,6}$. Recent detection of the early ultraviolet pulse of a peculiar subluminous SN Ia has been claimed as new evidence for the companion-ejecta interaction through the single-degenerate channel$^{7,8}$. Here, we report the discovery of a prominent but red optical flash at $\sim$ 0.5 days after the explosion of a SN Ia which shows hybrid features of different SN Ia sub-classes: a light curve typical of normal-brightness SNe Ia, but with strong titanium absorptions, commonly seen in the spectra of subluminous ones. We argue that the early flash of such a hybrid SN Ia is different from predictions of previously suggested scenarios such as the companion-ejecta interaction$^{8-10}$. Instead it can be naturally explained by a SN explosion triggered by a detonation of a thin helium shell either on a near-Chandrasekhar-mass white dwarf ($\gtrsim$ 1.3 M$_{\odot}$) with low-yield $^{56}$Ni or on a sub-Chandrasekhar-mass white dwarf ($\sim$ 1.0 M$_{\odot}$) merging with a less massive white dwarf. This finding provides compelling evidence that one branch of the previously proposed explosion models, the helium-ignition scenario, does exist in nature, and such a scenario may account for explosions of white dwarfs in a wider mass range in contrast to what was previously supposed$^{11-14}$.}

\end{spacing}


\subsection*{}

\begin{spacing}{2.0}

A faint optical transient was discovered on UT April 4.345, 2016 through the newly established high-cadence deep-imaging survey which is optimized for finding Type Ia Supernovae (SNe Ia) within a few days after explosion with the Subaru/Hyper Suprime-Cam (HSC)$^{15}$---``the \textbf{MU}lti-band \textbf{S}ubaru \textbf{S}urvey for \textbf{E}arly-phase \textbf{S}Ne Ia" (\textbf{MUSSES}). Close attention has been paid to one transient because its brightness increased by $\sim$ 6.3 times within one day of the first observation. We designated this fast-rising transient as MUSSES1604D (the official designation is SN~2016jhr)---the fourth early-phase SN candidate found in the April 2016 observing run of MUSSES.

\mbox{}

Figure 1 presents the observed $g$-, $r$-, $i$-band light curves of MUSSES1604D. The earliest photometry by Subaru/HSC indicates an apparent $g$-band magnitude of 25.14 $\pm$ 0.15 on April 4.345 (MJD 57482.345). One day later, MUSSES1604D brightened rapidly to $\sim$ 23.1 and 23.0 mag in the $g$ and $r$ bands, respectively. More surprisingly, the $g$-band observation on April 6 indicates that the transient ``paused" brightening from April 5, showing a plateau-like evolution lasting for $\sim$ 1 day. At the same time, the transient also slowed down in its rate of brightening in the $r$ band.

\mbox{}

Follow-up observations indicated that MUSSES1604D is a SN Ia with a $r$-band peak absolute magnitude of $\sim$ -19.1 on April 26. Adopting a host galaxy redshift $z$ of 0.11737, the rest-frame light curves in the $B$- and $V$-band absolute magnitudes from $\sim$ 4 days after the first observation are derived by applying a K-correction based on the best-fitting model with SALT2$^{16}$. Because of the peculiar flash at early time, K-correction for the flash-phase light curves is performed by simplified spectral-energy distributions, estimated from the early color information of MUSSES1604D (see Methods). The rest-frame $B$-band light curve shows a peak absolute magnitude of about -18.8 and $\Delta$$m_{15}$($B$) $\approx$ 1.0 mag, indicating a normal-brightness SN Ia$^{17}$.

\mbox{}

Color evolution within a few days after a SN explosion is crucial for identifying the early flash$^{8,10}$. In contrast to another peculiar early-flash SN Ia iPTF14atg with $B-V$ color evolution obtained only from $\sim$ 5 days after the discovery$^{7}$, the specific survey strategy of the MUSSES project enables us to obtain the color information of MUSSES1604D from 1 day after the first observation (Figure 2), which shows a slightly red $B-V$ color of about 0.2 mag at first, reddening further to about 0.5 mag in one day.

\mbox{}

The interaction of SN ejecta with a non-degenerate companion star$^{8,18,19}$ (``companion-ejecta interaction", CEI) or with dense circumstellar material$^{9,10}$ (``CSM-ejecta interaction") are popular scenarios to explain the early optical flash. In order to produce a prominent optical flash comparable to that of MUSSES1604D, either a companion with a very extended envelope or a large-scale CSM distribution is required. In the CEI scenario, a prominent flash generated from the inner, hot region of ejecta can be observed through the hole that is carved out by a red-giant companion$^{8,19}$. In the CSM-ejecta-interaction scenario, a more extended CSM distribution could generate a brighter flash but with longer diffusion time. Our best-fitting CEI model (Figures 2 \& 3) and previous simulations of both two scenarios$^{8,10,19}$ all indicate that the particular blue color evolution is inevitable when producing the early flash as bright as that of MUSSES1604D (Extended Data Figure 1), which is incompatible with the red and rapid early color evolution observed for MUSSES1604D.

\mbox{}

Peculiar spectral features have been discovered around the peak epoch (Figure 4). At first glance, the Si II $\lambda$6355 line, the W-shaped S II feature, and the Ca II H \& K absorptions are reminiscent of a normal SN Ia, while the weak Si II $\lambda$5972 line suggests a higher photospheric temperature than those of SNe Ia with similar luminosities. On the other hand, prominent absorption features such as the Ti II trough around 4150 $\AA$, usually attributed to low temperature, have been found at the same time, in contrast to the brightness indicated by the light curve. By inspecting near-maximum spectra of more than 800 non-subluminous SNe Ia, we found just three MUSSES1604D-like objects---SN~2006bt, SN~2007cq and SN~2012df (Extended Data Figures 2 \& 3), indicating the rarity of such hybrid SNe Ia.

\mbox{}

The peculiar spectral features and the early flash followed by a normal-brightness light curve observed for MUSSES1604D are incompatible with predictions of classical explosion mechanisms$^{20,21}$ through the hydrogen-accreting single degenerate channel, but suggested by a specific scenario in which the SN explosion is triggered by the He-shell detonation, so-called the double-detonation (DDet) scenario$^{12,13,22,23}$. In principle, a He-shell detonation not only generates a shock wave propagating toward the center of the white dwarf (WD) and ignites carbon burning near the center, but also allocates $^{56}$Ni and other radioactive isotopes such as $^{52}$Fe and $^{48}$Cr to the outermost layers where the optical depth is relatively low$^{12,23}$. Therefore energy deposited by decaying radioactive isotopes diffuses out and consequently results in a prominent flash in the first few days after the explosion (see Methods). Observationally, the plateau-like light curve enhancement can be observed with the day-cadence observations. At the same time, a significant amount of not only iron group elements but also intermediate mass elements such as Ti and Ca will be produced in the outermost layers$^{12,13,23}$. Vast numbers of absorption lines of these elements are very effective in blocking the flux in the blue part of the optical spectrum, thus leading to a relatively red $B-V$ color evolution in general. Indeed, although a substantial amount of He is left after the detonation, the expected spectrum would not show a trace of He in the optical wavelength$^{24}$. By assuming a progenitor star with a WD mass of 1.03 $M_{\odot}$ and a He-shell mass as low as $\sim$ 0.054 $M_{\odot}$ (as required to trigger the He detonation on the surface of a 1.03 $M_{\odot}$ WD$^{12,23}$), the prominent early flash, peculiar early color evolution and Ti II trough feature are reproduced simultaneously (Figures 2--4). Early-phase photometric behavior similar to that seen in our simulation has also been independently shown in a simulation of the sub-Chandrasekhar DDet model very recently$^{25}$, validating our simulation and interpretation.

\mbox{}

A potential issue in our simulation is the assumption of a sub-Chandrasekhar-mass WD with a thin He-shell. The amount of synthesized $^{56}$Ni is sensitive to the mass of the exploding WD and determines the peak luminosity$^{12,23}$. The DDet model requires a sub-Chandrasekhar-mass WD ($\sim$ 1 $M_{\odot}$) for the peak luminosity of MUSSES1604D. However, DDet happening on such a WD would lead to a fast-evolving $B$-band light curve, which is inconsistent with a much slower-evolving light curve observed for MUSSES1604D. In addition, the early flash resulting from the corresponding He mass of 0.054 $M_{\odot}$ is much brighter than that of MUSSES1604D. We suggest two alternative scenarios that also involve He detonation to solve this issue. A He-ignited violent merger$^{14}$ can easily trigger a detonation in a thin He shell, and could produce the light curve of MUSSES1604D, but by fine-tuning the configuration of the binary system. Whether core detonation can be triggered by the thin-He-shell detonation through the WD-WD merger is also an open question$^{26,27}$. Alternatively, the lower mass He can be detonated on the surface of a near-Chandrasekhar-mass WD, which provides a better and more straightforward account of the light curve and spectral features (Figures 2--4). Further investigation suggests that the best-fitting WD mass is in the range 1.28--1.38 $M_\odot$ but with a low-yield $^{56}$Ni compared with the prediction by DDet (see Methods). This finding suggests that there could be a mechanism to reduce the mass of $^{56}$Ni in the explosion triggered by the He detonation. For example, the shock wave generated by He detonation may trigger a deflagration rather than a detonation near the centre of the WD$^{28}$, because the high degeneracy pressure of a near-Chandrasekhar-mass WD would inhibit the formation of a shock wave as strong as that seen in a sub-Chandrasekhar-mass WD. Although the observed peculiarities of MUSSES1604D could be naturally explained by this scenario, it is not yet clear how a thin He shell is formed on such a massive WD during binary evolution.

\mbox{}

The discovery of MUSSES1604D indicates that the He-detonation-triggered scenario is also promising to explain early-flash SNe Ia in addition to other popular scenarios$^{8-10}$. The prominent optical excess and peculiar color evolution in the earliest phase together with absorptions due to Ti II ions in around-maximum spectra can be used as indicators of this scenario. The slow-evolving $B$-band light curve makes the classical sub-Chandrasekhar DDet model previously supposed$^{11,12}$ unlikely. Recent work shows that the sub-Chandrasekhar DDet scenario could explain a part of normal SNe Ia if only a negligible amount of He exists at the time of the He-shell detonation$^{29,30}$. Given that MUSSES1604D is best explained by a He shell that is thin but still more massive than required in the above scenario, it opens up a possibility that the He-detonation-triggered scenario would produce a range of observational counterparts, controlled by the masses of both the WD and the He shell. The discovery of MUSSES1604D thus provides the first observational calibration about the range and combination of these quantities realized in nature.

\end{spacing}

\vspace{66 pt}


\subsection*{} 

\noindent 1. Filippenko, A. V. Optical Spectra of Supernovae. \emph{Ann. Rev. Astron. Astrophys.} \textbf{35,} 309--355 (1997).\\

\noindent 2. Maoz, D., Mannucci, F. \& Nelemans, G. Observational Clues to the Progenitors of Type Ia Supernovae. \emph{Ann. Rev. Astron. Astrophys.} \textbf{52,} 107--170 (2014).\\

\noindent 3. Perlmutter, S., \emph{et al.} Measurements of $\Omega$ and $\Lambda$ from 42 High-Redshift Supernovae. \emph{Astrophys. J.} \textbf{517,} 565--586 (1999).\\

\noindent 4. Riess, A. G., \emph{et al.} Observational Evidence from Supernovae for an Accelerating Universe and a Cosmological Constant. \emph{Astron. J.} \textbf{116,} 1009--1038 (1998).\\

\noindent 5. Hillebrandt, W. \& Niemeyer, J. C. Type IA Supernova Explosion Models. \emph{Ann. Rev. Astron. Astrophys.} \textbf{38,} 191--230 (2000).\\

\noindent 6. Whelan, J. \& Iben, I., Jr. Binaries and Supernovae of Type I. \emph{Astrophys. J.} \textbf{186,} 1007--1014 (1973).\\

\noindent 7. Cao, Y. \emph{et al.} A strong ultraviolet pulse from a newborn type Ia supernova. \emph{Nature.} \textbf{521,} 328--331 (2015).\\

\noindent 8. Kasen, D. Seeing the Collision of a Supernova with Its Companion Star. \emph{Astrophys. J.} \textbf{708,} 1025--1031 (2010).\\

\noindent 9. Levanon, N., Soker, N. \& Garc\'{i}a-Berro, E. Constraining the double-degenerate scenario for Type Ia supernovae from merger ejected matter. \emph{Mon. Not. R. Astron. Soc.} \textbf{447,} 2803--2809 (2015).\\

\noindent 10. Piro, A. L. \& Morozova, V. S. Exploring the Potential Diversity of Early Type Ia Supernova Light Curves. \emph{Astrophys. J.} \textbf{826,} 96 (2016).\\

\noindent 11. Bildsten, L., Shen, K. J., Weinberg, N. N. \& Nelemans, G. Faint Thermonuclear Supernovae from AM Canum Venaticorum Binaries. \emph{Astrophys. J. Lett.} \textbf{662,} L95--L98 (2007).\\

\noindent 12. Fink, M., \emph{et al.} Double-detonation sub-Chandrasekhar supernovae: can minimum helium shell masses detonate the core? \emph{Astron. \& Astrophys.} \textbf{514,} 53 (2010).\\

\noindent 13. Woosley, S. E. \& Kasen, D. Sub-Chandrasekhar Mass Models for Supernovae. \emph{Astrophys. J. Lett.} \textbf{747,} 38 (2011).\\

\noindent 14. Pakmor, R., Kromer, M., Taubenberger, S. \& Springel, V. Helium-ignited Violent Mergers as a Unified Model for Normal and Rapidly Declining Type Ia Supernovae. \emph{Astrophys. J. Lett.} \textbf{770,} L8 (2013).\\

\noindent 15. Miyazaki, S., \emph{et al.} Hyper Suprime-Cam. \emph{Proc. SPIE Conf. Ser.} \textbf{8446,} 84460Z-1--84460Z-9 (2012).\\

\noindent 16. Guy, J., \emph{et al.} SALT2: using distant supernovae to improve the use of type Ia supernovae as distance indicators. \emph{Astron. \& Astrophys.} \textbf{466,} 11--21 (2007).\\

\noindent 17. Phillips, M. M. The absolute magnitudes of Type IA supernovae. \emph{Astrophys. J. Lett.} \textbf{413,} L105--L108 (1993).\\

\noindent 18. Pan, K., Ricker, P. M. \& Taam, R. E. Impact of Type Ia Supernova Ejecta on Binary Companions in the Single-degenerate Scenario. \emph{Astrophys. J.} \textbf{750,} 151 (2012).\\

\noindent 19. Kutsuna, M. \& Shigeyama, T. Revealing progenitors of type Ia supernovae from their light curves and spectra. \emph{Publ. Astron. Soc. Jap.} \textbf{67,} 54 (2015).\\

\noindent 20. Nomoto, K., Thielemann, F. -K. \& Yokoi, K. Accreting white dwarf models for type I supern. III. Carbon deflagration supernovae. \emph{Astrophys. J.} \textbf{286,} 644--658 (1984).\\

\noindent 21. Khokhlov, A. M. Delayed detonation model for type IA supernovae. \emph{Astron. \& Astrophys.} \textbf{245,} 114--128 (1991).\\

\noindent 22. Guillochon, J., Dan, M., Ramirez-Ruiz, E. \& Rosswog, S. Surface Detonations in Double Degenerate Binary Systems Triggered by Accretion Stream Instabilities. \emph{Astrophys. J. Lett.} \textbf{709,} L64--L69 (2010).\\

\noindent 23. Kromer, M., \emph{et al.} Double-detonation Sub-Chandrasekhar Supernovae: Synthetic Observables for Minimum Helium Shell Mass Models. \emph{Astrophys. J.} \textbf{719,} 1067--1082 (2010).\\

\noindent 24. Boyle, A., Sim, S. A., Hachinger, S. \& Kerzendorf, W. Helium in Double-Detonation Models of Type Ia Supernovae. \emph{Astron. \& Astrophys.} \textbf{599,} 46 (2017).\\

\noindent 25. Noebauer, U. M., \emph{et al.} Early light curves for Type Ia supernova explosion models. Preprint at (http://arxiv.org/abs/1706.03613) (2017).\\

\noindent 26. Shen, K. J. \& Bildsten, L. The Ignition of Carbon Detonations via Converging Shock Waves in White Dwarfs. \emph{Astrophys. J.} \textbf{785,} 61 (2014).\\

\noindent 27. Tanikawa, A., \emph{et al.} Hydrodynamical Evolution of Merging Carbon-Oxygen White Dwarfs: Their Pre-supernova Structure and Observational Counterparts. \emph{Astrophys. J.} \textbf{807,} 40 (2015).\\

\noindent 28. Nomoto, K., Sugimoto, D. \& Neo, S. Carbon Deflagration Supernova, an Alternative to Carbon Detonation. \emph{Astrophys Space Sci.} \textbf{39,} L37--L42 (1976).\\

\noindent 29. Blondin, S., Dessart, L., Hillier, D. J. \& Khokhlov. A. M. Evidence for sub-Chandrasekhar-mass progenitors of Type Ia supernovae at the faint end of the width-luminosity relation. \emph{Mon. Not. R. Astron. Soc.} \textbf{470,} 157--165 (2017).\\

\noindent 30. Shen, K. J., Kasen, D., Miles, B. J. \& Townsley, D. M. Sub-Chandrasekhar-mass white dwarf detonations revisited. Preprint at (http://arxiv.org/abs/1706.01898) (2017).\\

\noindent

\clearpage

\begin{spacing}{2.0}

\mbox{}

\noindent\textbf{Author Contributions}~~J.J. initiated the study, carried out analysis and wrote the manuscript as the principal investigator of the MUSSES project. M.D. contributed to the initiation of the MUSSES project, and assisted with manuscript preparation and analysis together with K.M. and T.S. K.M. and T.S. organized the efforts for theoretical interpretation with J.J. and M.D. K.M. investigated the He-detonation-triggered explosion models and conducted radiation-transfer calculations used to generate simulated light curves and spectra. T.S. developed and ran the radiation-transfer calculations used to generate simulated CEI-induced light curves. K.N. provided insights into the He-detonation-triggered explosion models and assisted with the analysis. N.Y., H.F. and S.M. are core software developers for HSC and are in charge of the HSC Subaru Strategic Program project. N.Y., N.T. and M.T. developed the HSC transient server for selecting real-time supernova candidates and contributed to Subaru/HSC observations and data reduction. T.M. contributed to the Subaru/HSC observation and to target-of-opportunity observations made with the 1.05-m Kiso Schmidt telescope. S.W.J. contributed SALT spectroscopy and data reduction. \v{Z}.I., A.J.C., P.Y., P.R.-L., N.S., F.P., D.B., J.M., L.W., M.D.S., D.J., P.A.M. and C.A. are core collaborators of the MUSSES project who are in charge of follow-up observations (including proposal preparations) with the following telescopes: 3.5-m ARC, 10.4-m GTC, 8.1-m VLT, 2.5-m NOT, 2.5-m INT and 2-m LT. All of the authors contributed to discussions.

\mbox{}

\noindent\textbf{Author Information}~~The authors declare no competing financial interests. Correspondence and requests for materials should be addressed to J.J. (email: yuzhoujiang@ioa.s.u-tokyo.ac.jp).

\mbox{}

\noindent\textbf{Acknowledgements}~~The authors thank S. C. Leung and M. Kokubo for helpful discussions. We also thank the staff at the Southern African Large Telescope, the Gemini-North telescope, the Nordic Optical Telescope, the Isaac Newton Telescope, the Liverpool Telescope and the Kiso Schmidt telescope for observations and people who carried out follow-up observations which were unfruitful owing to the weather. Simulations for the He detonation models were carried out on a Cray XC30 at the Center for Computational Astrophysics, National Astronomical Observatory of Japan. The work is partly supported by the World Premier International Research Center Initiative (WPI Initiative), MEXT, Japan, Grants-in-Aid for Scientific Research of JSPS (16H01087 and 26287029 for M.D. and J.J.; 26800100 and 17H02864 for K.M.; 16H06341, 16K05287, and 15H02082 for T.S.; 26400222, 16H02168, and 17K05382 for K.N.; 15H02075, 16H02183, and 17H06363 for M.T.; 15H05892 for S.M.) and the research grant program of the Toyota foundation (D11-R-0830). S.W.J. acknowledges support from the US National Science Foundation through award AST-1615455. M.D.S acknowledges generous support provided by the Danish Agency for Science and Technology and Innovation realized through a Sapere Aude Level 2 grant, the Instrument-center for Danish Astrophysics (IDA), and by a research grant (13261) from VILLUM FONDEN. The Hyper Suprime-Cam (HSC) collaboration includes the astronomical communities of Japan and Taiwan, and Princeton University. The HSC instrumentation and software were developed by the National Astronomical Observatory of Japan (NAOJ), the Kavli Institute for the Physics and Mathematics of the Universe (Kavli IPMU), the University of Tokyo, the High Energy Accelerator Research Organization (KEK), the Academia Sinica Institute for Astronomy and Astrophysics in Taiwan (ASIAA), and Princeton University. Funding was contributed by the FIRST program from Japanese Cabinet Office, the Ministry of Education, Culture, Sports, Science and Technology (MEXT), the Japan Society for the Promotion of Science (JSPS), Japan Science and Technology Agency (JST), the Toray Science Foundation, NAOJ, Kavli IPMU, KEK, ASIAA, and Princeton University. The Pan-STARRS1 Surveys (PS1) have been made possible through contributions of the Institute for Astronomy, the University of Hawaii, the Pan-STARRS Project Office, the Max-Planck Society and its participating institutes, the Max Planck Institute for Astronomy, Heidelberg and the Max Planck Institute for Extraterrestrial Physics, Garching, The Johns Hopkins University, Durham University, the University of Edinburgh, Queen's University Belfast, the Harvard-Smithsonian Center for Astrophysics, the Las Cumbres Observatory Global Telescope Network Incorporated, the National Central University of Taiwan, the Space Telescope Science Institute, the National Aeronautics and Space Administration under Grant No. NNX08AR22G issued through the Planetary Science Division of the NASA Science Mission Directorate, the National Science Foundation under Grant No. AST-1238877, the University of Maryland, and Eotvos Lorand University (ELTE). This paper makes use of software developed for the Large Synoptic Survey Telescope. We thank the LSST Project for making their code available as free software at http://dm.lsst.org. This work is also based on observations obtained at the Gemini Observatory (program: GN-2016A-DD-7), which is operated by the Association of Universities for Research in Astronomy, Inc., under a cooperative agreement with the NSF on behalf of the Gemini partnership: the National Science Foundation (United States), the National Research Council (Canada), CONICYT (Chile), Ministerio de Ciencia, Tecnolog\'{i}a e Innovaci\'{o}n Productiva (Argentina), and Minist\'{e}rio da Ci\^{e}ncia, Tecnologia e Inova\c{c}\~{a}o (Brazil).

\end{spacing}

\clearpage

\setlength\intextsep{-11pt}
\begin{wrapfigure}{R}{1\textwidth}
\setlength{\belowcaptionskip}{-5pt} 
\includegraphics[width=1.0\textwidth]{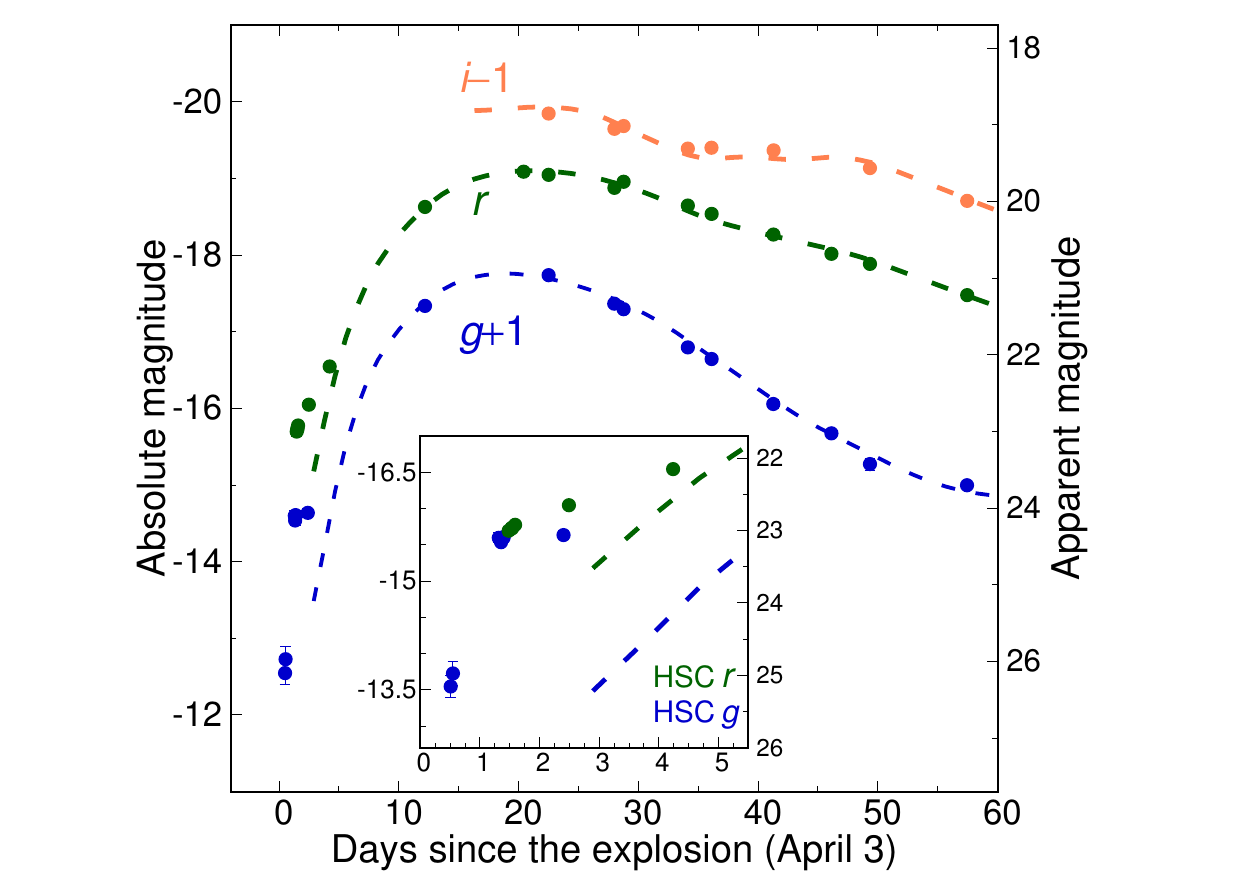}
\centering
\vspace{-5 pt}
\caption*{\textbf{Figure 1: The multi-band light curve of MUSSES1604D.} Photometry in $g$, $r$ and $i$ bands (observer-frame) are in the AB system. Error bars denote 1-$\sigma$ uncertainties. Dashed lines are best-fitting light curves derived from the non-early photometry (t $\gtrsim$ 12 days) with SALT2$^{16}$. The explosion epoch is estimated by adopting a classical $t$$^2$ fireball model for the early-flash phase (see Methods). The inset zooms in on the early-phase multi-band light curve by Subaru/HSC, which shows that the brightening in $g$-band ``paused" after the second-night observation.}

\end{wrapfigure}

\clearpage

\setlength\intextsep{-11pt}
\begin{wrapfigure}{R}{1\textwidth}
\setlength{\belowcaptionskip}{10pt} 
\includegraphics[width=1\textwidth]{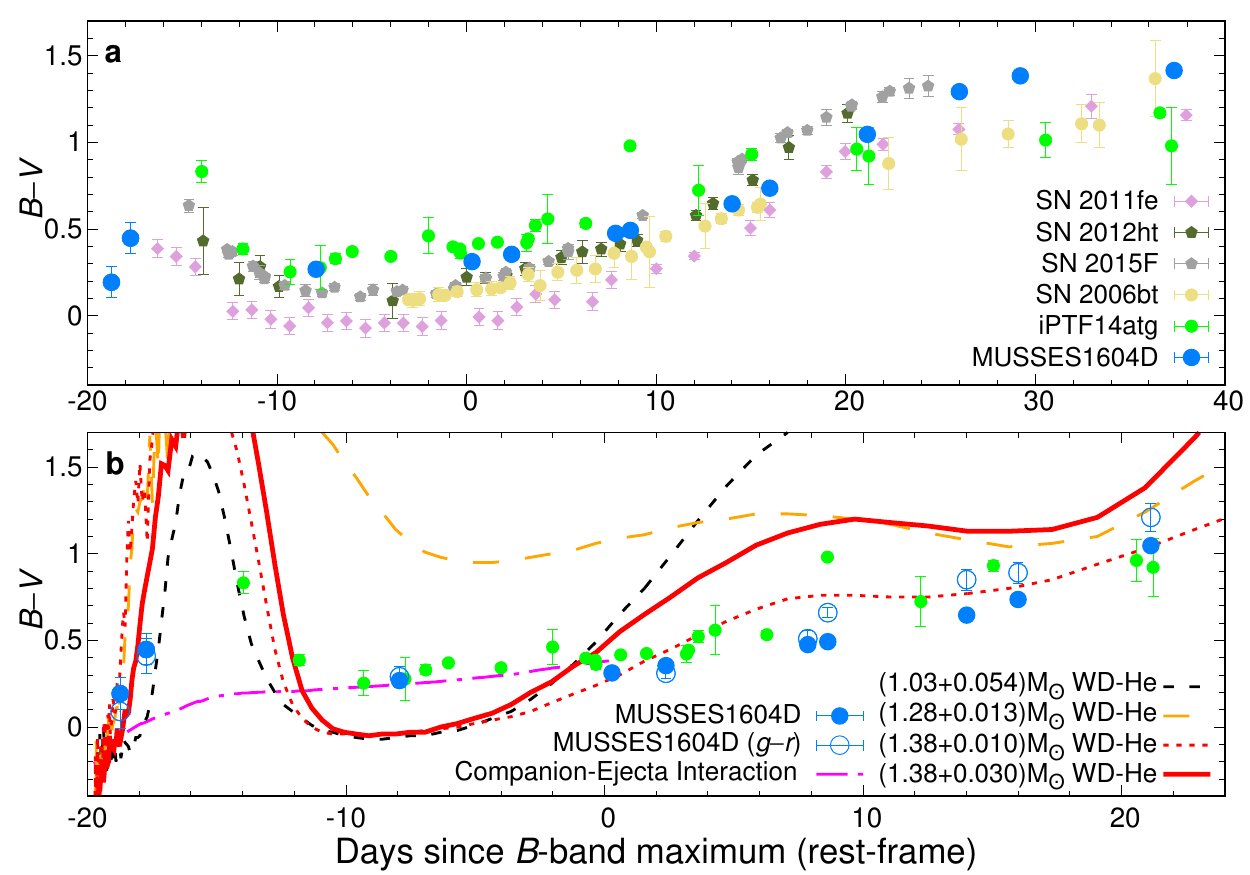}
\centering
\vspace{0 pt}
\caption*{\textbf{Figure 2: Comparative analysis of MUSSES1604D color evolution.} The upper panel presents $B-V$ color evolution of MUSSES1604D, iPTF14atg (early-flash), SN~2006bt (MUSSES1604D-like), SN~2012ht (transitional), SN~2015F and SN~2011fe (normal-brightness). The lower panel shows the color evolution predicted by CEI, He-detonation model for the sub-Chandrasekhar-mass WD and the newly proposed He-detonation models for the near-Chandrasekhar-mass WD under different He-shell mass assumptions. The $B$-band maximum occurred about 20 days after the explosion. As the bandpass difference between the rest-frame $B$/$V$ band and the observer-frame $g$/$r$ band is inconspicuous at $z$ $\sim$ 0.1, the observed $g-r$ color evolution is provided for reference. Error bars represent 1-$\sigma$ uncertainties.}

\end{wrapfigure}

\clearpage

\begin{figure}[tbp]
  \begin{subfigure}[b]{0.58\textwidth}
    \includegraphics[width=\textwidth]{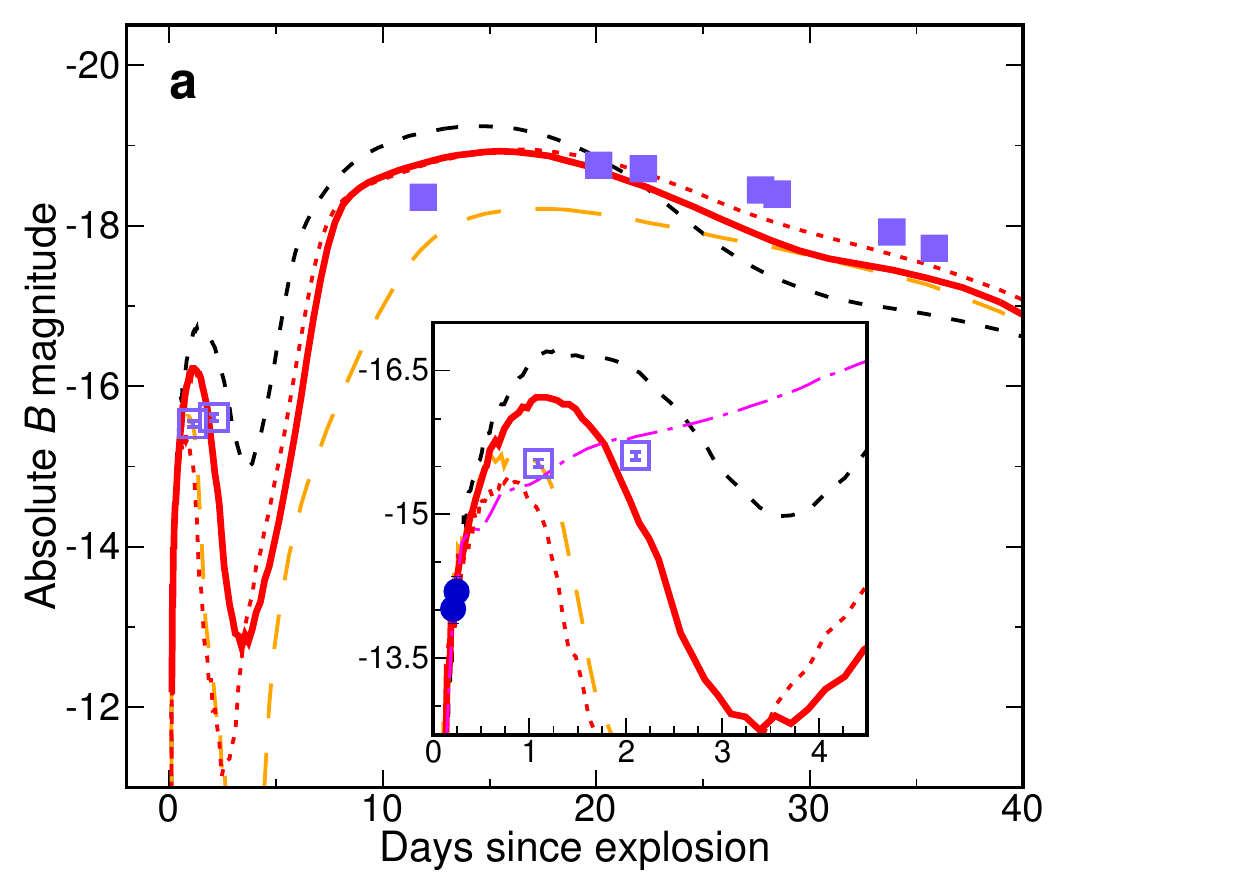}
  \end{subfigure}
  \hspace{-45pt}
  \begin{subfigure}[b]{0.58\textwidth}
    \includegraphics[width=\textwidth]{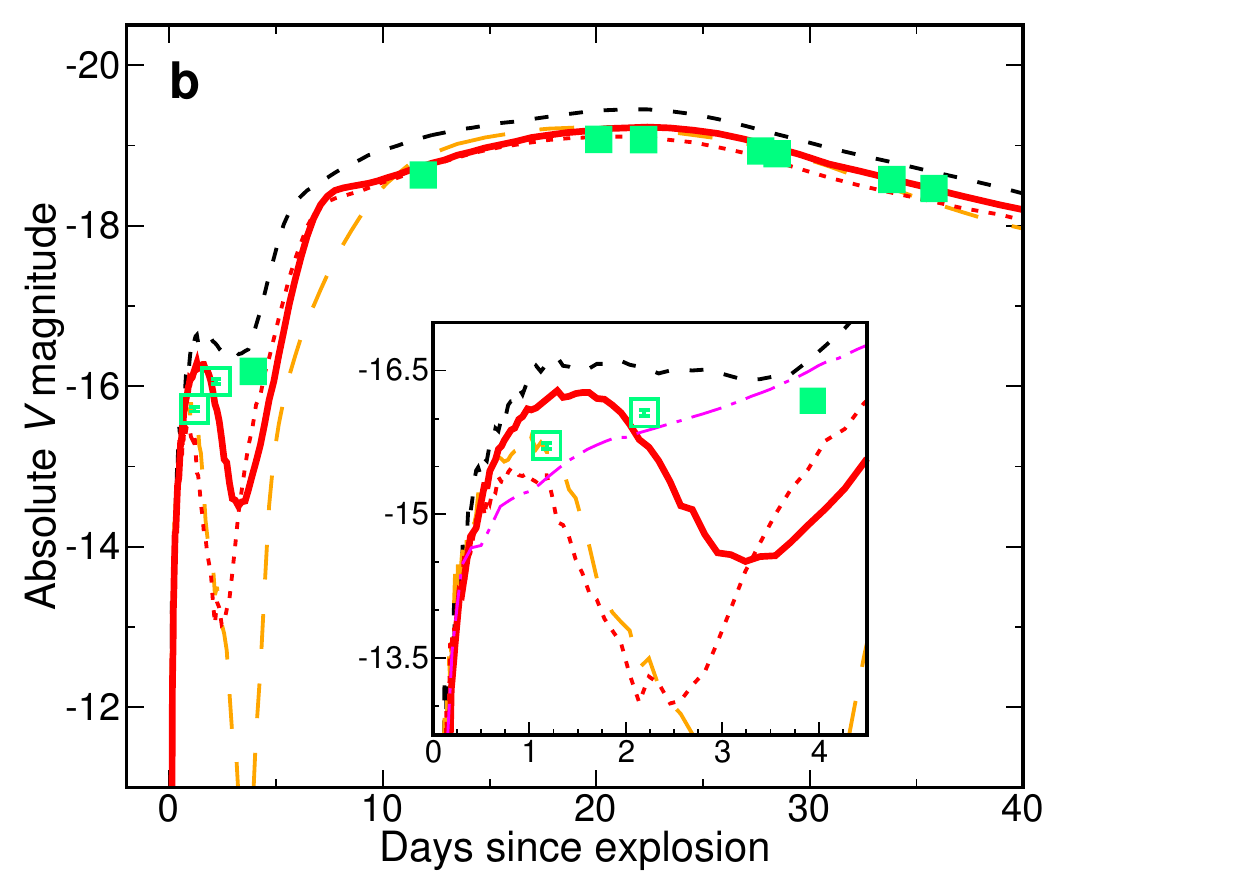}
  \end{subfigure}

\vspace{20 pt}
\caption*{\textbf{Figure 3: Rest-frame $B$- and $V$-band light curves of MUSSES1604D and simulations.} K-corrections in the flash (open squares) and the post-flash phase (filled squares) are carried out with different methods. Each panel includes He-detonation models for sub-Chandrasekhar-mass WD (1.03 $M_{\odot}$ WD + 0.054 $M_{\odot}$ He-shell; black dashed line) and massive WD (1.28 $M_{\odot}$ WD + 0.013 $M_{\odot}$ He-shell, orange long-dashed line; 1.38 $M_{\odot}$ WD + 0.01 $M_{\odot}$ He-shell, red dotted line; 1.38 $M_{\odot}$ WD + 0.03 $M_{\odot}$ He-shell, red solid line) conditions. The inset zooms in on the flash phase and also includes our best-fitting CEI model assuming a 1.05 $M_{\odot}$ red-giant companion (magenta dashed-dotted line). The first-night $g$-band data (blue circles) are included in panel \textbf{a}. The explosion epoch shown here is shifted (+0.3 days) from that estimated by the classical $t^2$ model (Figure 1) within the uncertainty from the simulations. Error bars denote 1-$\sigma$ uncertainties.}

\end{figure}

\clearpage

\setlength\intextsep{-11pt}
\begin{wrapfigure}{R}{1\textwidth}
\setlength{\belowcaptionskip}{0pt} 
\includegraphics[width=1\textwidth]{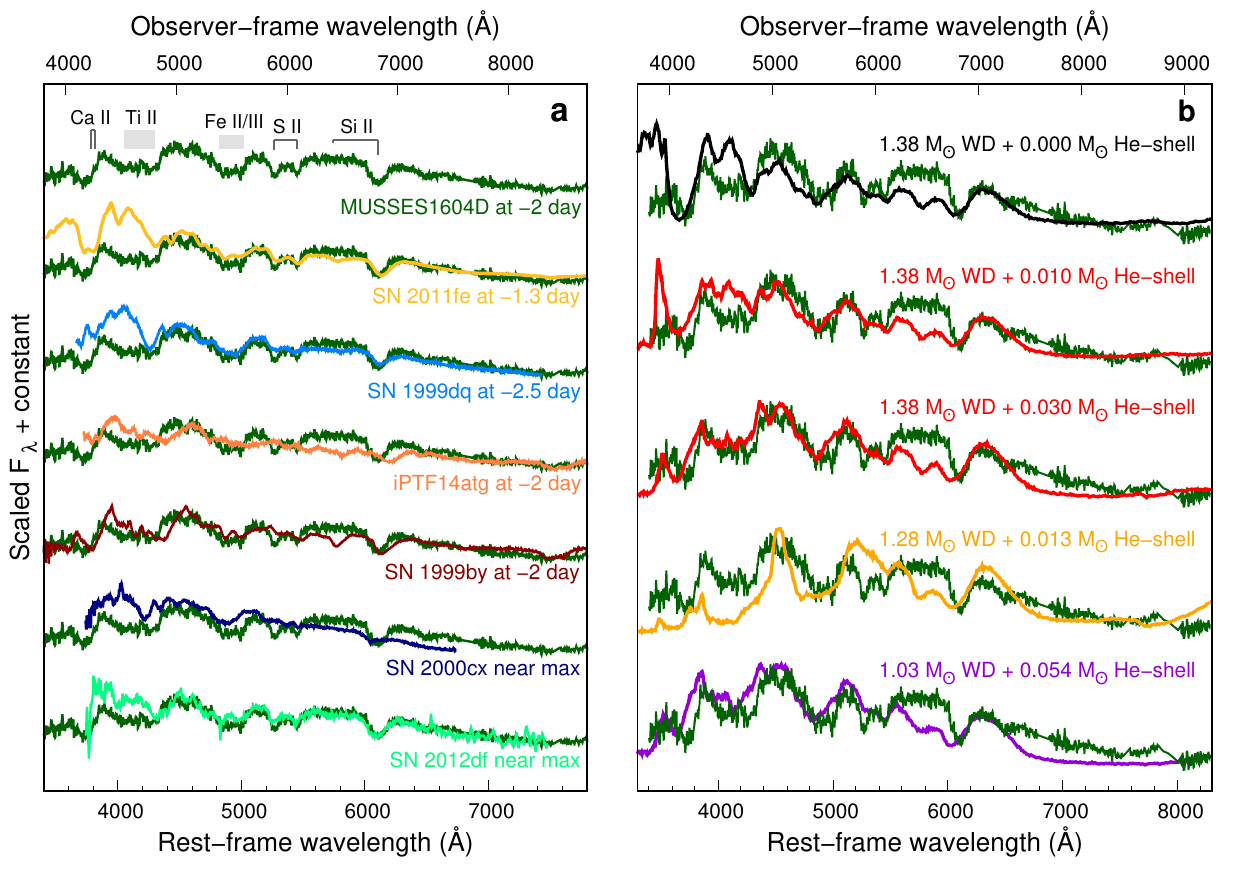}
\centering
\vspace{0pt}
\caption*{\textbf{Figure 4: An around-maximum spectral comparison of MUSSES1604D, other observed SNe Ia of different types, and models.} In panel \textbf{a}, the spectrum of MUSSES1604D taken 2 days before the $B$-band maximum by Southern African Large Telescope (SALT) is compared with that of SN~2011fe (normal), SN~1999dq and SN 2000cx (shallow-silicon), SN~1999by (subluminous), iPTF14atg (early-flash) and SN~2012df (MUSSES1604D-like) at a similar epoch. Major absorption features are labeled on the spectrum of MUSSES1604D. In panel \textbf{b}, simulated spectra of the classical W7 deflagration model (top), the newly proposed He-detonation models with different assumed He-shell masses (middle three), and the classical double-detonation model for a sub-Chandrasekhar-mass WD (bottom) are compared with the MUSSES1604D spectrum (dark green) at the same epoch (two days before the $B$-band maximum).}

\end{wrapfigure}

\clearpage


\subsection*{Methods}

\vspace{30pt}

\begin{spacing}{2.0}

\noindent\textbf{\uppercase\expandafter{\romannumeral1}. The Handbook for MUSSES1604D}

\mbox{}

\textbf{The MUSSES project and discovery of MUSSES1604D} The Subaru Hyper Suprime-Cam$^{15}$ (HSC) is a new-generation, wide-field camera which started to serve as a facility instrument of the 8.2-m Subaru telescope from 2014. With a total of 116 CCDs, a single HSC pointing covers 1.8 square degrees and reaches to a $g$-band limiting magnitude (5-$\sigma$) of about 26.5 mag with exposure time of 300 s.

\mbox{}

The \textbf{MU}lti-band \textbf{S}ubaru \textbf{S}urvey for \textbf{E}arly-phase \textbf{S}Ne Ia (\textbf{MUSSES}) is a newly established project which aims to systematically investigate the photometric and spectroscopic behavior of SNe Ia within a few days of their explosions (hereafter early-phase SNe Ia, ESNe Ia) with Subaru/HSC and other 1--10 m class telescopes around the world. In every semester, we plan to carry out 1--2 observing runs and each of them includes two stages: the Subaru/HSC survey (2--3 nights) and follow-up observations. In the survey stage, Subaru/HSC observes over 100 square degrees of sky with a $g$-band limiting magnitude of 26.0 (5-$\sigma$) every night for finding ESNe Ia and obtaining their multi-band light curve information. Using the HSC transient pipeline and newly employed machine-learning classifiers, we are able to carry out real-time candidate selection during the survey and trigger photometric/spectroscopic follow-ups within one day after the Subaru observation. Because of the fast brightening of ESNe Ia, photometric follow-up observations can be conducted well with 1--4 m telescopes. The strategy of MUSSES gives a very large photometric dynamic range, enabling us to observe ESNe Ia even to redshift $z$ $\sim$0.3.

\mbox{}

In order to make the best use of the Subaru time, the MUSSES observing run in April 2016 adopted a specific survey mode which combines both HSC Subaru Strategic Program (HSC SSP$^{31}$, 1-night $g$-band observation, from UT April 4.17 to UT April 4.67) and open-use observation (1.5-nights $g$- and $r$-band observation, from UT April 5.17 to 5.67 and April 6.43 to 6.67 respectively).

\mbox{}

The supernova MUSSES1604D (official designation: SN~2016jhr) was discovered on UT April 4.345, 2016 at $\alpha$(J2000) = 12h18m19s.85 and $\delta$(J2000) = +00$^\circ$15'17.38" with a $g$-band magnitude of 25.14 mag upon discovery (Extended Data Figure 4). This was the fourth ESN candidate found in the April observing run. MUSSES1604D was located about 5.8" (to the southwest) from the host-galaxy center. The redshift of the host galaxy is 0.11737 $\pm$ 0.00001 according to the SDSS (Data Release 12)$^{32}$. With cosmological parameters H$_{0}$ = 70 km s$^{-1}$ Mpc$^{-1}$, $\Omega$$_{m}$ = 0.30, $\Omega$$_{\Lambda}$ = 0.70 and $\Omega$$_{\nu}$ = 0.00, we calculate a luminosity distance of 546.5 megaparsecs and a distance modulus of 38.69 mag for MUSSES1604D.

\mbox{}

\textbf{The host galaxy} The red color with a visible H$\alpha$ emission feature suggests that the host galaxy of MUSSES1604D is a star-forming early-type galaxy$^{33}$. Further analysis of the SDSS photometry and spectroscopy shows that the stellar mass is $3-7 \times 10^{10}$ $M_{\odot}$, which is also consistent with an early-type galaxy, e.g. an S0 galaxy.

\mbox{}

\textbf{Follow-up observations} Our scheduled early follow-up observations at La Palma island and Apache Point Observatory were lost owing to poor weather conditions. By HSC SSP $r$-band observation conducted at two days after our Subaru/HSC observations, we successfully took another $r$-band image of MUSSES1604D, which provides a crucial constraint on the timescale of the early-flash. Multi-band follow-up observations with the 8.0-m Gemini-North telescope, the 3.5-m Astrophysical Research Consortium (ARC) telescope, the 2.5-m Nordic Optical Telescope (NOT), the 2.5-m Isaac Newton Telescope (INT), the 2-m Liverpool Telescope (LT) and the 1.05-m Kiso Schmidt telescope have been conducted from about -8 days to +40 days after the $B$-band maximum. For the spectroscopic observations, we triggered the 9.2-m SALT and the 8.0-m Gemini-North telescope at specific epochs to get spectral evolution from about -2 days to one month after the $B$-band maximum (Extended Data Figure 2).

\mbox{}

\textbf{Data reduction and photometric calibration} As MUSSES1604D resides at the edge of the host galaxy, contamination from the host is negligible except for the photometry of the earliest Subaru/HSC observation. The morphology of the host galaxy indicates a symmetric S0 galaxy. We thus built the host template with GALFIT$^{34,35}$ and performed the standard point spread function (PSF) photometry with the IRAF DAOPHOT package$^{36}$ on host-subtracted images. The photometry has been tested by subtracting the SN from the original image, using an artificial PSF star with the derived photometric magnitude. The average flux of the residual region is comparable with the surrounding region and well below the photometric error of the discovery image by Subaru/HSC (the 1-$\sigma$ photometric error is 0.15 mag). PSF photometry is performed on host-subtracted images for all follow-up observations as well. The photometry is then calibrated to the standard SDSS photometric system by adopting a color term correction based on field stars$^{37}$. For spectroscopic data reduction, all data were reduced with standard routines in IRAF.

\mbox{}

\noindent\textbf{\uppercase\expandafter{\romannumeral2}. Light curve fitting and K-correction}

\mbox{}

Considering the limited understanding of spectral features during the early optical flash phase of MUSSES1604D, we adopted different methods to derive the rest-frame light curves at flash and post-flash phases respectively. For the post-flash light curves, we firstly fit the observed light curves by applying the SALT2 model of SNe Ia spectrophotometric evolution which is built using a large data set including light curves and spectra of both nearby and distant SNe Ia$^{16}$. After light curve fitting, K-correction is performed to get the rest-frame $B$- and $V$-band light curves according to the best-fitting spectral sequence model of MUSSES1604D with SNCosmo$^{38}$. For the light curve in the flash phase (within 5 days after the explosion), we applied the color-based K-correction with a pseudo power-law spectral energy distribution (SED) function $f$($\nu$) = $k$$\nu$$^\alpha$, where $\nu$ is the frequency of the light, $k$ and $\alpha$ are parameters derived by the early color information of MUSSES1604D. Considering there is no indication of Na I D absorption lines in any of our spectra ($S/N$ $\sim$ 18 per resolution element near the wavelength of Na I D lines for the around-maximum spectrum) and the supernova was located far away from the center of an S0-type host, we only take into account the Galactic extinction given by E($B-V$)$_{MW}$ = 0.0263 mag (SFD, 1998$^{39}$). The rest-frame $B$- and $V$-band light curves are shown in Extended Data Figure 3.

\mbox{}

The K-corrected rest-frame light curve of MUSSES1604D indicates a $B$-band peak absolute magnitude of -18.8, but with $\Delta$$m_{15}$($B$) $\approx$ 1.0 mag, corresponding to a slow-evolving normal-brightness SN Ia according to the Phillips relation$^{17}$. The $V$-band light curve of MUSSES1604D is consistent with typical normal-brightness SNe Ia, such as SN~2011fe. All photometric data in observer and rest frames are listed in Extended Data Table 1.

\mbox{}

\noindent\textbf{\uppercase\expandafter{\romannumeral3}. Explanations for the peculiarities of MUSSES1604D} 

\mbox{}

The ``peculiarities" of MUSSES1604D mainly include: \textbf{1)} a prominent optical flash with peculiar color evolution at very early time; \textbf{2)} the red $B-V$ color evolution in general; \textbf{3)} a normal-brightness SN Ia with prominent Ti II absorptions in the around-maximum spectrum; \textbf{4)} a slow-evolving $B$-band light curve. In this section, we compare different scenarios which may account for such peculiarities, and find the best solution.

\mbox{}

\textbf{The companion-ejecta interaction.} We performed two-dimensional axisymmetric radiation hydrodynamic simulations of the explosions of a WD with a Chandrasekhar mass in binary systems to obtain light curves and spectra resulting from collisions between the ejecta and the companion star (Kutsuna and Shigeyama's (K-S) CEI models$^{19,40}$). The ejecta are described by the W7 model$^{20}$. The best-fitting light curves presented in Figure 3 are the outcomes expected from an explosion in a binary system with a separation of 2.5$\times10^{13}$ cm when we observe this event from the companion side. The companion star is a red giant with a mass of 1.05 $M_{\odot}$ (the core mass is 0.45 $M_{\odot}$) and a radius of 8.9$\times10^{12}$ cm, filling the Roche lobe. The initial mass of the companion was assumed to be 1.50 $M_{\odot}$. Although the CEI-induced early flash could be prominent in this condition, we cannot reproduce the early light curves and $B-V$ color evolution of MUSSES1604D because a strong but long-lasting flash will be produced after interacting with a red giant which has a more extended envelope$^{8,19}$. For the spectral peculiarity (Figure 4), the prominent Ti II lines also contradict the predictions of typical explosion models through the hydrogen-accreting single degenerate channel$^{5,20,21}$.

\mbox{}

Further comparisons of early-phase light curves with both Kasen's (K10) and K-S CEI models$^{8,19}$ are presented in panels \textbf{a}--\textbf{c} of Extended Data Figure 1. Note that K10 predicts a brighter early flash than K-S models because it assumes instantaneous thermalization in the shocked matter while K-S models approximately take into account thermalization processes between shocked matter and radiation (cooling of shocked matter by bremsstrahlung). As K10 noticed, the assumption of instantaneous thermalization tends to underestimate the energies of photons and also results in overestimating the emissivity from shocked matter. Therefore, K10 models produce a prominent flash even with a low-mass main-sequence companion while K-S models can only marginally produce a comparable early flash with a red-giant companion, and produce an even fainter early flash with a main-sequence companion. Despite the different assumptions in two CEI models, with an early flash as bright as that of MUSSES1604D, both K10 and K-S models predict blue color of $B-V$ $\lesssim$ 0.1 in the first 4 days after the explosion, which is incompatible with the observations of MUSSES1604D.

\mbox{}

\textbf{The CSM-ejecta interaction.} In the double-degenerate progenitor scenario where a SN Ia is generated from the merger of two WDs, a considerable amount of material from the disrupted secondary WD may get pushed out to a large radius$^{41,42}$ and possibly result in an early ultraviolet/optical flash due to the interaction with the ejecta$^{9,10}$. The strong early light curve enhancement seen in MUSSES1604D requires a very extended CSM distribution$^{10,43}$. Regardless of the physical possibility of reaching the CSM distribution under their assumptions, interactions with more extended CSM not only strengthen the early flash but also increase the diffusion time, resulting in a bluer and longer flash phase$^{10}$. Panels \textbf{d}--\textbf{f} of Extended Data Figure 1 show the early light curves and color evolution predicted by CSM-ejecta interaction models. To produce a flash with a brightness comparable to that of MUSSES1604D, blue and slow color evolution is inevitable in these models, even after fine-tuning the CSM scale and the $^{56}$Ni distribution of the inner ejecta. Therefore, the CSM-ejecta interaction cannot explain the prominent early flash and the rapid, red $B-V$ color evolution observed for MUSSES1604D.

\mbox{}

\textbf{The He-detonation-triggered scenarios.} Another scenario is the SN Ia explosion triggered by the detonation of the He layer. The He detonation generates radioactive materials as the nucleosynthesis ash. For example, He detonation on the surface of a Chandrasekhar-mass WD would leave $^{56}$Ni as a main energy source in this layer with the mass fraction ($X_{\rm 56Ni}$) reaching to $\sim$ 20\% (see below). The diffusion time$^{44}$ of optical photons through this He layer is estimated to be $\sim$ 2 days $\times (\kappa/0.2 {\rm cm}^2 {\rm g}^{-1})^{0.5} (M_{\rm He}/0.02 M_{\odot})^{0.5} (V_{\rm He}/20,000 {\rm km} {\rm s}^{-1})^{-0.5}$. Here, a subscript ``He" is used for quantities related to the He layer, and the He as a dominant element in the layer is assumed to be fully ionized. $M_{\rm He}$ and $V_{\rm He}$ are the mass and velocity of the He layer respectively, and$\kappa$ is the opacity. The decay power at $\sim$ 2 days from the $^{56}$Ni in the He layer is estimated to be $\sim 2.5 \times 10^{41}$ erg s$^{-1}$ $(X_{\rm 56Ni}/0.2) \times (M_{\rm He}/0.02 M_{\odot})$. Therefore, the radioactivity in the He-detonation ash is predicted to produce a prompt flash lasting for a few days with the peak bolometric magnitude of $\sim$ -16, assuming the He mass $M_{\rm He}$ $\sim$ 0.03 $M_{\odot}$. This scenario roughly explains the nature of the early flash found for MUSSES1604D. For the sub-Chandrasekhar WD, the abundance in the He ash is dominated by the other radioactive isotopes, $^{52}$Fe and $^{48}$Cr, and they power the early flash. Still, a similar argument as above applies.

\mbox{}

The synthetic light curves and spectra expected from the He-shell detonation models are simulated as follows (Figures 2--4). We constructed a series of toy one-dimensional models which mimic the results of DDet hydrodynamic simulations$^{13,23}$. The density structure is assumed to be exponential in velocity space, where the kinetic energy is specified by the energy generation for the assumed burned composition structure. A stratified structure in the composition and a uniform abundance pattern in each layer are assumed, where the distribution of the burning products is set to represent the DDet models$^{23}$.

\mbox{}

The model structures are shown in Extended Data Figures 5 and 6. Our sub-Chandrasekhar model and Chandrasekhar model are similar to a typical DDet model and the W7 model, respectively, in the mass coordinate. For the Chandrasekhar WD model, we replaced part of the $^{56}$Ni-rich region with a Si-rich region, leading to a more centrally confined structure than in the W7 model. Note that we assume a stable Fe/Ni region in the core of the Chandrasekhar model, the mass of which is taken to be $\sim$ 0.2 $M_{\odot}$ similar to that in the W7 model. For each model, we run multi-frequency and time-dependent Monte-Carlo radiation transfer calculations$^{40}$, which were updated to include radioactive energy input from the decay chains of $^{52}$Fe/Mn/Cr and $^{48}$Cr/V/Ti together with $^{56}$Ni/Co/Fe. The code assumes LTE for the ionization, which is generally believed to be a good approximation in the early phase. For example, in the W7 model, LTE and NLTE simulations yield indistinguishable light curves (except for the $U$-band) until $\sim$ 25 days after the explosion, corresponding to $\sim$ 5 days after the $B$-maximum$^{45,46}$. In addition, we do not include the non-thermal excitation of He; generally He absorption lines in optical wavelength are invisible for the DDet models even with this effect$^{24}$.

\mbox{}

The peculiar early light curve and color evolution as well as the strong Ti II absorptions for MUSSES1604D can be naturally reproduced by the DDet scenario, as shown by the model for the 1.03 $M_{\odot}$ WD with 0.054 $M_{\odot}$ He (ash) layer in Figures 2--4. We note that the idea that the DDet model predicts an early flash by the radioactive decay of He ash was independently proposed by Noebauer et al.$^{25}$ in work posted on arXiv after we submitted this paper. Their model, which is qualitatively similar to our sub-Chandrasekhar model, leads to the early flash and the color evolution in the first few days after an explosion, powered by the decay of $^{52}$Fe and $^{48}$Cr, as is similar to our model prediction. However, their model lacks around/post-maximum light-curve and spectral information, therefore further comparison with our model is not possible.

\mbox{}

Although the sub-Chandrasekhar DDet model can explain most of peculiarities of MUSSES-1604D, it has prominent defects in the resulting fast evolution of the simulated $B$-band light curve (see also refs$^{13,23}$). Note that the fast decline of $B$-band light curve predicted by this classical DDet scenario of the sub-Chandrasekhar-mass WD happens from $\sim$ 17 days after the explosion, and the magnitude becomes $\sim$ 1.4 mag fainter than the peak at t $\sim$ 25 days. The difference between the LTE and NLTE treatments in the first 25 days after the explosion is too small to account for such abnormal light curve evolution$^{45,46}$, and thus it is unlikely that the LTE assumption accounts for this discrepancy. Another issue is that the quantity of radioactive isotopes in this model (0.054 $M_{\odot}$ of the He layer as a minimal He shell for the He detonation) produces a stronger flash than that of MUSSES1604D, suggesting that the observationally required mass of the He layer is lower. For further investigation of the classical DDet scenario, we ran a grid of models spanning WD masses from $\sim$ 0.9 to 1.4 $M_{\odot}$, but all the models predicted very fast evolution in the $B$-band light curve and/or too-bright peak luminosity.

\mbox{}

Indeed, this fast evolution in the $B$-band light curve has been recognized as one of the issues in the (sub-Chandrasekhar) DDet model$^{13}$, because the Fe-peak and Ti/Cr in the He ash should start blocking the photons in the bluer bands once the temperature decreases after the maximum light, and this argument is not sensitive to the LTE or NLTE treatment. It has been shown that this problem could be remedied if the mass of the He layer is much smaller than that required by the classical DDet Model so as not to provide a large opacity$^{29,30,47}$, partly based on an idea that such a small amount of He ($<$ 0.01 $M_\odot$) would lead to detonation when a substantial fraction of carbon is mixed in the He layer$^{48}$. We have also confirmed from our model sequence that the light curves of sub-Chandrasekhar DDet models are indeed roughly consistent with normal (but relatively faint and fast-evolving) SNe Ia, once the He layer is removed. However, this scenario would not explain MUSSES1604D, as we do see prominent early flash and signatures of the He ash in the maximum spectrum.

\mbox{}

To remedy the abnormal fast-evolution issue in the classical DDet scenario, we investigated additional models in which we allow that the relation between the WD mass and the final $^{56}$Ni production expected in DDet is not necessarily fulfilled. By altering the WD mass, $^{56}$Ni mass, and the He mass, the most straightforward choice we found is shown in Figures 2--4, where the models with 1.38 $M_{\odot}$ WD, 0.01--0.03 $M_{\odot}$ He-ash layer, and 0.43 $M_{\odot}$ of $^{56}$Ni are presented. Additionally we investigated the model with 1.28 $M_\odot$ WD, 0.013 $M_\odot$ He-ash layer, and 0.44 $M_{\odot}$ of $^{56}$Ni. While such a relatively less massive WD model can also give a slow-evolving light curve, the pre-maximum $B-V$ color turns out to be too red. Therefore, we constrain the acceptable WD mass range between 1.28 and 1.38 $M_\odot$. In addition, an even better consistency of the light curves and color evolution from $\sim$ 5 days after the $B$-band maximum could be expected for our preferred model (1.38 $M_{\odot}$ WD + 0.03 $M_{\odot}$ He-shell) once the NLTE effects were taken into account$^{45,46}$.

\mbox{}

From these analyses, we suggest two scenarios that involve He detonation. First is the He-ignited violent merger scenario$^{14}$. In this case, the primary WD mass should still be $\sim$ 1 $M_{\odot}$ to produce the required peak luminosity. The accretion stream of He during the merging process may trigger a detonation even if the He mass is low$^{14,27}$. If the secondary white dwarf is swept up by the ejecta, this would explain the slow-evolving light curve. However, there are two drawbacks to this scenario: (1) it is uncertain whether the core detonation can be triggered by the thin-He-shell detonation$^{26,27}$; and (2) it will involve fine-tuning of the merging configuration (e.g., the masses of the WDs) to reproduce the observational features of MUSSES1604D.

\mbox{}

The second scenario is the He detonation on the surface of a nearly-Chandrasekhar-mass WD, as is motivated by our light curve and spectral models which reproduce the observational results quite well simply by assuming a standard Chandrasekhar-mass WD without fine-tuning. The amount of He mass is also consistent in this picture to trigger the detonation there. The evolutionary track of this binary system towards the He detonation on the surface of a WD more massive than 1.3 M$_{\odot}$ has been never discussed in the literature. Further investigations are needed to explore whether this scenario can be realized or not. Another drawback is that in the classical DDet scenario it will produce too much $^{56}$Ni through core detonation, resulting in an over-luminous SN Ia. Still, the fact that this simple model explains all the main features of MUSSES1604D is striking, indicating that it is unlikely to be a mere coincidence. This would suggest that there could be a mechanism to reduce the mass of $^{56}$Ni as compared to the classical DDet picture.

\mbox{}

More realistic light curve and spectra might be realized if one takes into account the possible viewing-angle effect related to both the violent merger scenario and the He-ignited near-Chandrasekhar-WD scenario. Our one-dimensional models only address angle-averaged behavior. The Ti/Fe absorptions will be stronger than our one-dimensional prediction if the line of sight is to intersect a region of the He ash. The initial light-curve enhancement would also be dependent on the viewing angle, but this effect would be much less prominent than in the absorption.

\mbox{}

Another issue is that the Si and S features of MUSSES1604D are not very well reproduced. In general, these features are qualitatively well explained but obtaining quantitatively good fits is an issue even with sophisticated NLTE modeling$^{49}$. We find that these features are also sensitive to detailed composition structure even in one-dimensional simulations. Providing detailed fitting for these features is beyond the scope of this study, as these features are theoretically more uncertain than the features we have analyzed in this paper.

\mbox{}

\noindent\textbf{\uppercase\expandafter{\romannumeral4}. The explosion epoch of MUSSES1604D}

\mbox{}

Extrapolating the explosion epoch of a SN Ia based on the $^{56}\textrm{Ni}$-powered light curve is controversial because a considerable ``dark phase" between the explosion and the radioactive decay from SN ejecta may exist for some SNe Ia$^{50-53}$. For example, the best-observed SN Ia so far, SN~2011fe likely has a one-day dark phase though it was discovered at the brightness of $\sim$ 1/1000 of its peak brightness$^{52,54}$. More stringent restrictions on the explosion epoch require not only deep-imaging observations but also specific radiation mechanisms at early times to light up the dark phase$^{55}$. Thanks to the deep imaging capability of Subaru/HSC and the early flash of MUSSES1604D, the explosion time of MUSSES1604D can be pinpointed.

\mbox{}

In He-detonation-triggered scenarios, the early optical flash is produced immediately from the radioactive decay at the surface of the SN ejecta, which is the earliest optical emission except for the almost non-detectable cooling emission from the shock-heated WD soon after the SN shock breakout$^{50,56}$. Thus, MUSSES1604D was discovered at an earlier phase than any previously discovered SN Ia. Given an effectively negligible dark phase before the early flash, we adopt the classical $t$$^2$ fireball model (where $t$ is the time since the explosion) for the rising phase of the early flash, assuming that neither the photospheric temperature nor the velocity changes significantly in estimating the explosion epoch of MUSSES1604D. The result indicates that the first observation of MUSSES1604D is at $\sim$ 0.51$\small{^{+0.08}_{-0.06}}$ days after the SN explosion. According to the best-fitting light curves derived from the post-flash multi-band photometry (dashed lines in Figure 1), the $g$-band magnitude reaches the same level of our first observation (25.14 mag) at t $\sim$ 3 days. This may imply that a non-negligible dark phase exists for non-early-flash SNe Ia.

\mbox{}

\noindent\textbf{\uppercase\expandafter{\romannumeral5}. 
MUSSES1604D-like SNe Ia and their rarity}

\mbox{}

The rate of occurrence of He-detonation-triggered SNe Ia can be constrained by estimating the fraction of MUSSES1604D-like SNe Ia. By inspecting over 1,000 SNe Ia from normal to various different subtypes which have at least one good spectrum from about -6 to +12 days after their $B$-band maximum through published resources and open SN databases$^{57,58}$, three MUSSES1604D-like SNe Ia (without early-phase observations) have been found. The screening criteria and detailed properties of the MUSSES1604D-like SNe Ia are listed in Extended Data Table 2. In addition to three normal-brightness SNe Ia (-19.4 $\lesssim$ M$_{B}$ $\lesssim$ -18.7) we mentioned here, some subluminous SNe Ia also show good similarities to MUSSES1604D (e.g. 02es-like SNe Ia, PTF10ops and SN~2010lp, which also have slow-evolving light curve and similar spectral features to MUSSES1604D$^{59,60}$). However, due to insufficient information to classify these subluminous objects conclusively, the discussion here focuses on the best three MUSSES1604D-like SNe Ia, namely SN~2006bt, SN~2007cq, and SN~2012df$^{61-64}$.

\mbox{}

A normal-brightness SN Ia, SN~2006bt shows good similarity with MUSSES1604D in both light curve and spectral features except for the shallow Si II $\lambda$5972 absorption seen in MUSSES1604D. Because there is no Na I D feature in the spectra of SN~2006bt and the SN is far away from the center of an S0/a host galaxy, the absolute magnitude is shown in Extended Data Table 2 without taking into account the host extinction. Well-organized follow-up observations for SN~2006bt indicate pre-maximum Ti II absorptions, a slow-evolving $B$-band light curve and similar $B-V$ color evolution to MUSSES1604D.

\mbox{}

SN~2007cq is classified as another MUSSES1604D-like SN Ia. In particular, the pre-maximum spectroscopy of SN~2007cq shows prominent Ti II absorptions from about 6 days before the $B$-band maximum, which is consistent with the prediction of the He-detonation models$^{12,23}$. Note that SN~2007cq shows shallower intermediate element absorption features and bluer color than MUSSES1604D, which could be attributed to a larger amount of $^{56}$Ni generated from the core explosion for SN~2007cq.

\mbox{}

SN~2012df was located at the edge of an S0-like galaxy. The spectrum was taken near its brightness peak with an unfiltered absolute magnitude of $\sim$ -18.9 (without extinction correction). Despite the limited observational information for SN~2012df, high spectral similarity between two SNe Ia has been found at a similar epoch (Figure 4). Therefore we classify SN~2012df as a MUSSES1604D-like SN Ia. Comparisons of the spectral evolution and light curves of MUSSES1604D-like SNe Ia are presented in Extended Data Figures 2 and 3 respectively.

\mbox{}

To obtain a conservative estimate of the event rate of MUSSES1604D-like SNe Ia, we eliminated all subluminous objects, even though some of them may have the same origin$^{59,60}$. Statistically there are 4 MUSSES1604D-like objects (including MUSSES1604D) out of $\sim$ 800 SNe Ia with $B$-band peak absolute magnitude $\lesssim$ -18.7, corresponding to a fraction of MUSSES1604D-like SNe Ia of $\sim$ 0.5\%.

\mbox{}

The traditional SN Ia classification which is mainly based on SN brightness and spectral features will classify MUSSES1604D and iPTF14atg into two peculiar subtypes, even though both have strong early light curve enhancements, slow-evolving light curves, prominent Ti II absorptions, and similar color evolution and host environments$^{7,55}$, implying that they might be intrinsically connected$^{65}$. However, whether iPTF14atg is also triggered by the He-shell detonation is an open question because the Ti II absorptions and red color of subluminous SNe Ia in the post-flash phase can be attributed to the low temperature of ejecta, and the lack of early color information prevents us from further comparisons with MUSSES1604D at the flash phase. It is worth noting that the earliest $B-V$ color of iPTF14atg at $\sim$ 5 days after the explosion is probably too red to be explained by CEI or CSM-ejecta interaction, but is in line with the predictions of He-detonation models (Figure 2). As a reference for the future work, in Extended Data Table 2, we list MUSSES1604D-like and iPTF14atg-like candidates selected from different SN Ia branches$^{59,60,66-68}$. Similarities among these objects may suggest intrinsic connections between a number of SNe Ia of different subtypes.

\mbox{}

\noindent \textbf{Code availability.} The post-flash light curve fitting and K-correction are carried out with the SALT2 model and SNCosmo, which are available at http://supernovae.in2p3.fr/salt/doku.php \& https://sncosmo.readthedocs.io/en/v1.5.x/ respectively. We have not made publicly available the code for the companion-ejecta interaction (CEI) models or the radiation-transfer code used for He-detonation simulations, because they are not prepared for the open-use. Instead, the simulated light curves and spectra for the He-detonation models shown in this paper are available upon request.

\mbox{}

\noindent \textbf{Data availability.} The Source Data for Figures 1, 3 and 4 are available in the online version of the paper. Photometric and spectroscopic data will also be made publicly available on WISeREP3 (http://wiserep.weizmann.ac.il/).

\mbox{}

\end{spacing}

\vspace{-20 pt}

\noindent 31. Miyazaki, S., \emph{et al.} Wide-field imaging with Hyper Suprime-Cam: Cosmology and Galaxy Evolution, A Strategic Survey Proposal for the Subaru Telescope. http://hsc.mtk.nao.ac.jp/ssp/wp-content/uploads/2016/05/hsc\_ssp\_rv\_jan13.pdf (2014).\\

\noindent 32. Bolton, A. S., \emph{et al.} Spectral Classification and Redshift Measurement for the SDSS-III Baryon Oscillation Spectroscopic Survey. Astron. J. \textbf{144,} 144 (2012).\\

\noindent 33. Shimasaku, K., \emph{et al.} Statistical Properties of Bright Galaxies in the Sloan Digital Sky Survey Photometric System. \emph{Astron. J.} \textbf{122,} 1238--1250 (2001).\\

\noindent 34. Peng, C. Y., Ho, L. C., Impey, C. D. \&  Rix, H. -W. Detailed Decomposition of Galaxy Images. II. Beyond Axisymmetric Models. \emph{Astron. J.} \textbf{139,} 2097--2129 (2010).\\

\noindent 35. The GALFIT software can be downloaded at \\ https://users.obs.carnegiescience.edu/peng/work/galfit/galfit.html.\\

\noindent 36. Stetson, P. B. DAOPHOT: A Computer Program for Crowded-Field Stellar Photometry. \emph{Publ. Astron. Soc. Pac.} \textbf{99,} 191--222 (1987).\\

\noindent 37. Doi, M., \emph{et al.} Photometric Response Functions of the Sloan Digital Sky Survey Imager. \emph{Astron. J.} \textbf{139,} 1628--1648 (2010).\\

\noindent 38. SNCosmo is available at https://sncosmo.readthedocs.io/en/v1.5.x/.\\

\noindent 39. Schlegel, D. J., Finkbeiner, D. P. \& Davis, M. Maps of Dust Infrared Emission for Use in Estimation of Reddening and Cosmic Microwave Background Radiation Foregrounds. \emph{Astrophys. J.} \textbf{500,} 525--553 (1998).\\

\noindent 40. Maeda, K., Kutsuna, M. \& Shigeyama, T. Signatures of a Companion Star in Type Ia Supernovae. \emph{Astrophys. J.} \textbf{794,} 37 (2014).\\

\noindent 41. Fryer, C. L., \emph{et al.} Spectra of Type Ia Supernovae from Double Degenerate Mergers. \emph{Astrophys. J.} \textbf{725,} 296--308 (2010).\\

\noindent 42. Shen, K. J., Bildsten, L., Kasen, D. \& Quataert, E. The Long-term Evolution of Double White Dwarf Mergers. \emph{Astrophys. J.} \textbf{748,} 35 (2012).\\

\noindent 43. Levanon, N. \& Soker, N. Early UV emission from disk-originated matter (DOM) in type Ia supernovae in the double degenerate scenario. \emph{Mon. Not. R. Astron. Soc.} \textbf{470,} 2510--2516 (2017).\\

\noindent 44. Arnett, D. Supernovae and Nucleosynthesis: An Investigation of the History of Matter from the Big Bang to the Present. Princeton University Press (1996).\\

\noindent 45. Kasen, D., Thomas, R. C. \& Nugent, P. Time-dependent Monte Carlo Radiative Transfer Calculations for Three-dimensional Supernova Spectra, Light Curves, and Polarization. \emph{Astrophys. J.} \textbf{651,} 366--380 (2006).\\

\noindent 46. Kromer, M. \& Sim, S. A. Time-dependent three-dimensional spectrum synthesis for Type Ia supernovae. \emph{Mon. Not. R. Astron. Soc.} \textbf{398,} 1809--1826 (2009).\\

\noindent 47. Sim, S. A.,\emph{et al.} Detonations in Sub-Chandrasekhar-mass C+O White Dwarfs. \emph{Astrophys. J. Lett.} \textbf{714,} L52--L57 (2010).\\ 

\noindent 48. Shen, K. J., \& Moore, K. The Initiation and Propagation of Helium Detonations in White Dwarf Envelopes. \emph{Astrophys. J.} \textbf{797,} 46 (2014).\\ 

\noindent 49. Nugent, P., Phillips, M., Baron, E., Branch, D. \& Hauschildt, P. Evidence for a Spectroscopic Sequence among Type 1a Supernovae. \emph{Astrophys. J. Lett.} \textbf{455,} L147--L150 (1995).\\

\noindent 50. Piro, A. L., \& Nakar, E. What can we Learn from the Rising Light Curves of Radioactively Powered Supernovae? \emph{Astrophys. J.} \textbf{769,} 67 (2013).\\

\noindent 51. Piro, A. L., \& Nakar, E. Constraints on Shallow $^{56}$Ni from the Early Light Curves of Type Ia Supernovae. \emph{Astrophys. J.} \textbf{784,} 85 (2014).\\

\noindent 52. Mazzali, P. A., \emph{et al.} Hubble Space Telescope spectra of the Type Ia supernova SN 2011fe: a tail of low-density, high-velocity material with Z $<$ Z$\odot$. \emph{Mon. Not. R. Astron. Soc.} \textbf{439,} 1959--1979 (2014).\\

\noindent 53. Zheng, W., \emph{et al.} Estimating the First-light Time of the Type Ia Supernova 2014J in M82. \emph{Astrophys. J. Lett.} \textbf{783,} L24 (2014).\\

\noindent 54. Nugent, P. E., \emph{et al.} Supernova SN 2011fe from an exploding carbon-oxygen white dwarf star. \emph{Nature.} \textbf{480,} 344--347 (2011).\\

\noindent 55. Cao, Y., \emph{et al.} SN2002es-like Supernovae from Different Viewing Angles. \emph{Astrophys. J.} \textbf{832,} 86 (2016).\\

\noindent 56. Piro, A. L., Chang, P. \& Weinberg, N. N. Shock Breakout from Type Ia Supernova. \emph{Astrophys. J.} \textbf{708,} 598--604 (2010).\\

\noindent 57. Yaron, O. \& Gal-Yam, A. WISeREP--An Interactive Supernova Data Repository. \emph{Publ. Astron. Soc. Pac.} \textbf{124,} 668--681 (2012).\\

\noindent 58. Guillochon, J., Parrent, J., Kelley, L. Z. \&  Margutti, R. An Open Catalog for Supernova Data. \emph{Astrophys. J.} \textbf{835,} 64 (2017).\\

\noindent 59. Maguire, K., \emph{et al.} PTF10ops -- a subluminous, normal-width light curve Type Ia supernova in the middle of nowhere. \emph{Mon. Not. R. Astron. Soc.} \textbf{418,} 747--758 (2011).\\

\noindent 60. Kromer, M., \emph{et al.} SN 2010lp--a Type Ia Supernova from a Violent Merger of Two Carbon-Oxygen White Dwarfs. \emph{Astrophys. J. Lett.} \textbf{778,} L18 (2013).\\

\noindent 61. Foley, R. J., \emph{et al.} SN 2006bt: A Perplexing, Troublesome, and Possibly Misleading Type Ia Supernova. \emph{Astrophys. J.} \textbf{708,} 1748--1759 (2010).\\

\noindent 62. Ganeshalingam, M., \emph{et al.} Results of the Lick Observatory Supernova Search Follow-up Photometry Program: BVRI Light Curves of 165 Type Ia Supernovae. \emph{Astrophys. J. Suppl.} \textbf{190,} 418--448 (2010).\\

\noindent 63. Scalzo, R., \emph{et al.} Type Ia supernova bolometric light curves and ejected mass estimates from the Nearby Supernova Factory. \emph{Mon. Not. R. Astron. Soc.} \textbf{440,} 1498--1518 (2014).\\

\noindent 64. Ciabattari, F., \emph{et al.} Supernova 2012df = Psn J17481875+5218023. \emph{Central Bureau Electronic Telegrams.} No. 3161 (2012).\\

\noindent 65. Kromer, M., \emph{et al.} The peculiar Type Ia supernova iPTF14atg: Chandrasekhar-mass explosion or violent merger? \emph{Mon. Not. R. Astron. Soc.} \textbf{459,} 4428--4439 (2016).\\

\noindent 66. Ganeshalingam, M., \emph{et al.} The Low-velocity, Rapidly Fading Type Ia Supernova 2002es. \emph{Astrophys. J.} \textbf{751,} 142 (2012).\\

\noindent 67. Li, W., \emph{et al.} SN 2002cx: The Most Peculiar Known Type Ia Supernova. \emph{Publ. Astron. Soc. Pac.} \textbf{115,} 453--473 (2003).\\

\noindent 68. Foley, R. J., \emph{et al.} Type Iax Supernovae: A New Class of Stellar Explosion. \emph{Astrophys. J.} \textbf{767,} 57 (2013).

\setlength\intextsep{-11pt}
\begin{wrapfigure}{R}{1\textwidth}
\setlength{\belowcaptionskip}{-5pt} 
\includegraphics[width=1.0\textwidth]{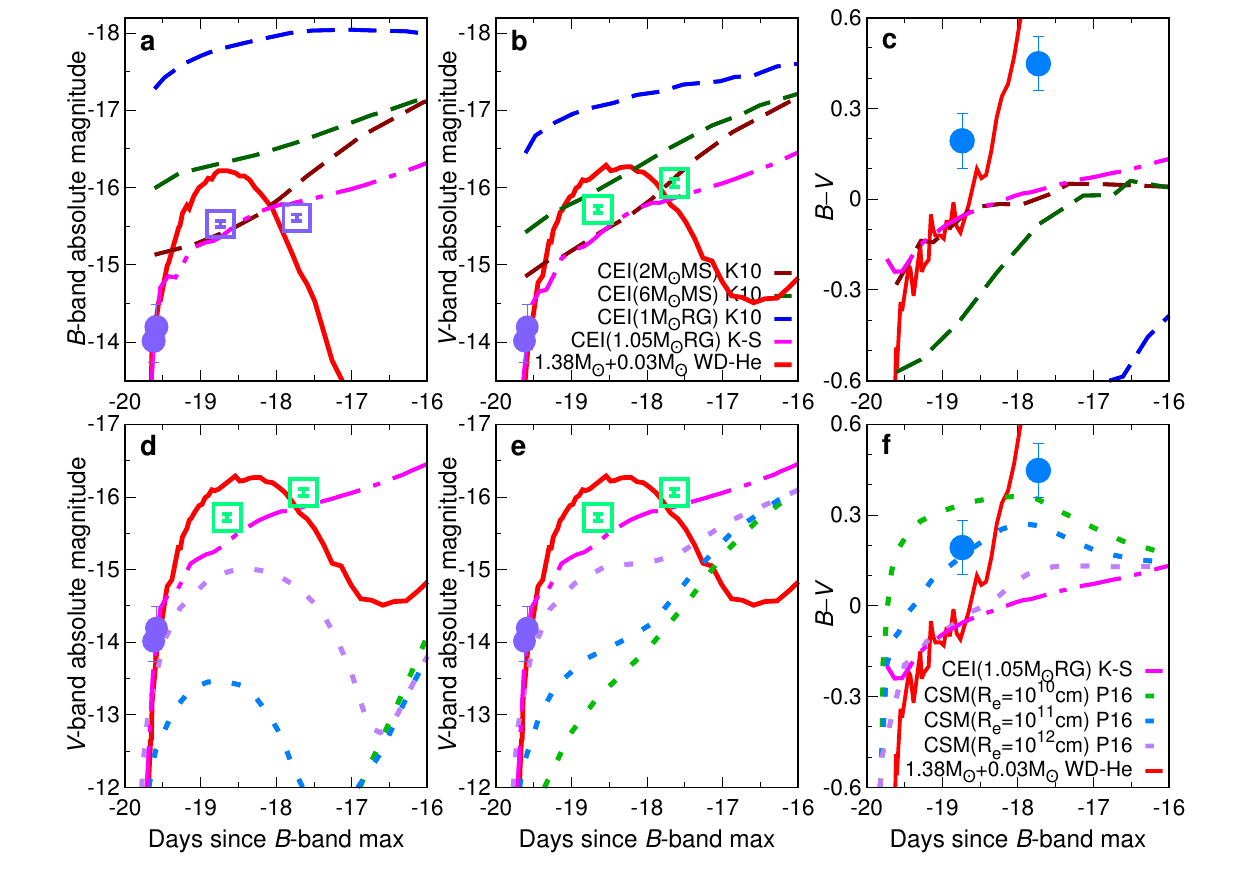}
\centering
\vspace{-20 pt}

\caption*{\textbf{Extended Data Figure 1: Comparison of MUSSES1604D observations and different model simulations at flash phase.} Symbols for MUSSES1604D data are the same as those in Figures 1--3 and the results from our best-fitting He-detonation model (1.38 $M_{\odot}$ WD + 0.03 $M_{\odot}$ He-shell, red solid lines) are shown in each panel. Panels \textbf{a-c} present early $B$-band (\textbf{a}) and $V$-band (\textbf{b}) light curves and $B-V$ color evolution (\textbf{c}) generated by different CEI simulations observed from the companion side. Dashed lines correspond to the K10 models with different binary-system compositions (MS, main-sequence star; RG, red-giant star)$^8$. The magenta dashed-dotted line denotes our best-fitting K-S CEI model$^{19}$. Although an early flash as bright as that of MUSSES1604D could be produced with specific CEI models, the predicted color is very blue at the CEI-flash phase. Panels \textbf{d} and \textbf{e} are $V$-band light curves simulated by the CSM-ejecta interaction with deep (\textbf{d}) and shallow (\textbf{e}) $^{56}$Ni distribution for the inner ejecta (Piro \& Morozova, P16$^{10}$). Dotted lines correspond to an external mass of M$_e$ = 0.3 $M_{\odot}$ with different outer radii R$_e$. Panel \textbf{f} is the color evolution under the same assumptions as in \textbf{e}. Similar to CEI models, combinations of early light curves and color evolution predicted by the CSM-ejecta interaction are different from the observed features of MUSSES1604D.}

\end{wrapfigure}

\clearpage

\setlength\intextsep{-11pt}
\begin{wrapfigure}{R}{1\textwidth}
\setlength{\belowcaptionskip}{10pt} 
\includegraphics[width=1\textwidth]{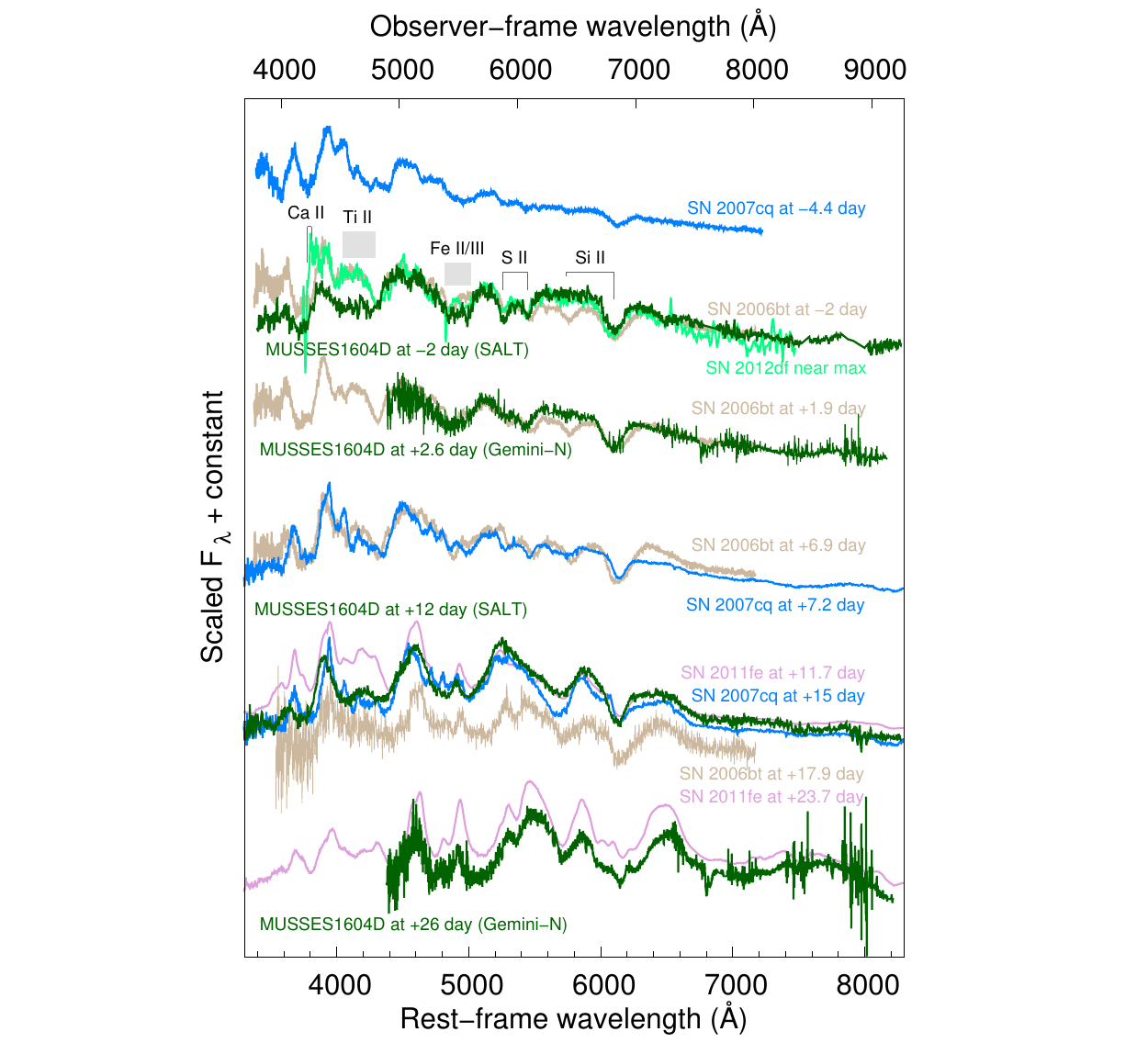}
\centering
\vspace{0 pt}
\caption*{\textbf{Extended Data Figure 2: Spectral evolution of MUSSES1604D and analogues.} Spectra for MUSSES1604D (dark green) are compared with those of the analogous SNe Ia SN~2006bt, SN~2007cq and SN~2012df at similar epochs. Late-phase spectra of SN~2011fe are included for reference. SALT/RSS follow-up observations were carried out -2 and 12 days after the $B$-band maximum and the other two spectra were taken by Gemini-N/GMOS 3 and 26 days after the $B$-band maximum.}
\end{wrapfigure}

\clearpage

\setlength\intextsep{-11pt}
\begin{wrapfigure}{R}{1\textwidth}
\setlength{\belowcaptionskip}{-5pt} 
\includegraphics[width=1.0\textwidth]{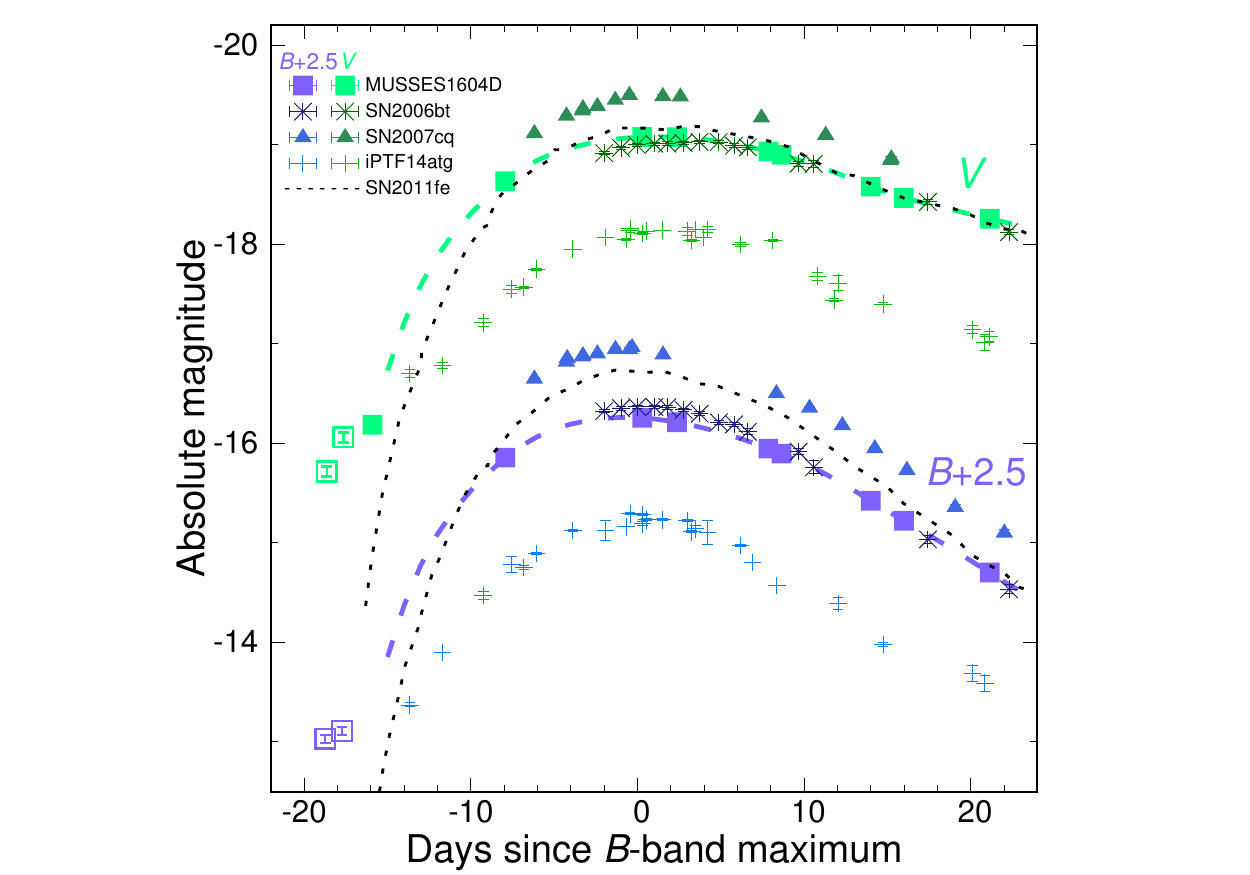}
\centering
\vspace{-5 pt}
\caption*{\textbf{Extended Data Figure 3: Rest-frame $B$- and $V$-band light curves for MUSSES1604D and other SNe Ia.} K-corrections in flash (open squares) and post-flash phase (filled squares with dashed lines) of MUSSES1604D were carried out with different methods (see Methods). An excellent light curve match is shown for MUSSES1604D, SN~2006bt and SN~2007cq. Another peculiar early-flash SN Ia iPTF14atg also shows similar light curves though its brightness is $\sim$ 1 magnitude fainter than MUSSES1604D. Light curves of a normal SN Ia, SN~2011fe (black dotted lines) are provided for reference. Magnitudes shown here are in the Vega system and the error bars denote 1-$\sigma$ uncertainties.}
\end{wrapfigure}

\clearpage

\setlength\intextsep{-11pt}
\begin{wrapfigure}{R}{1\textwidth}
\setlength{\belowcaptionskip}{10pt} 
\includegraphics[width=1\textwidth]{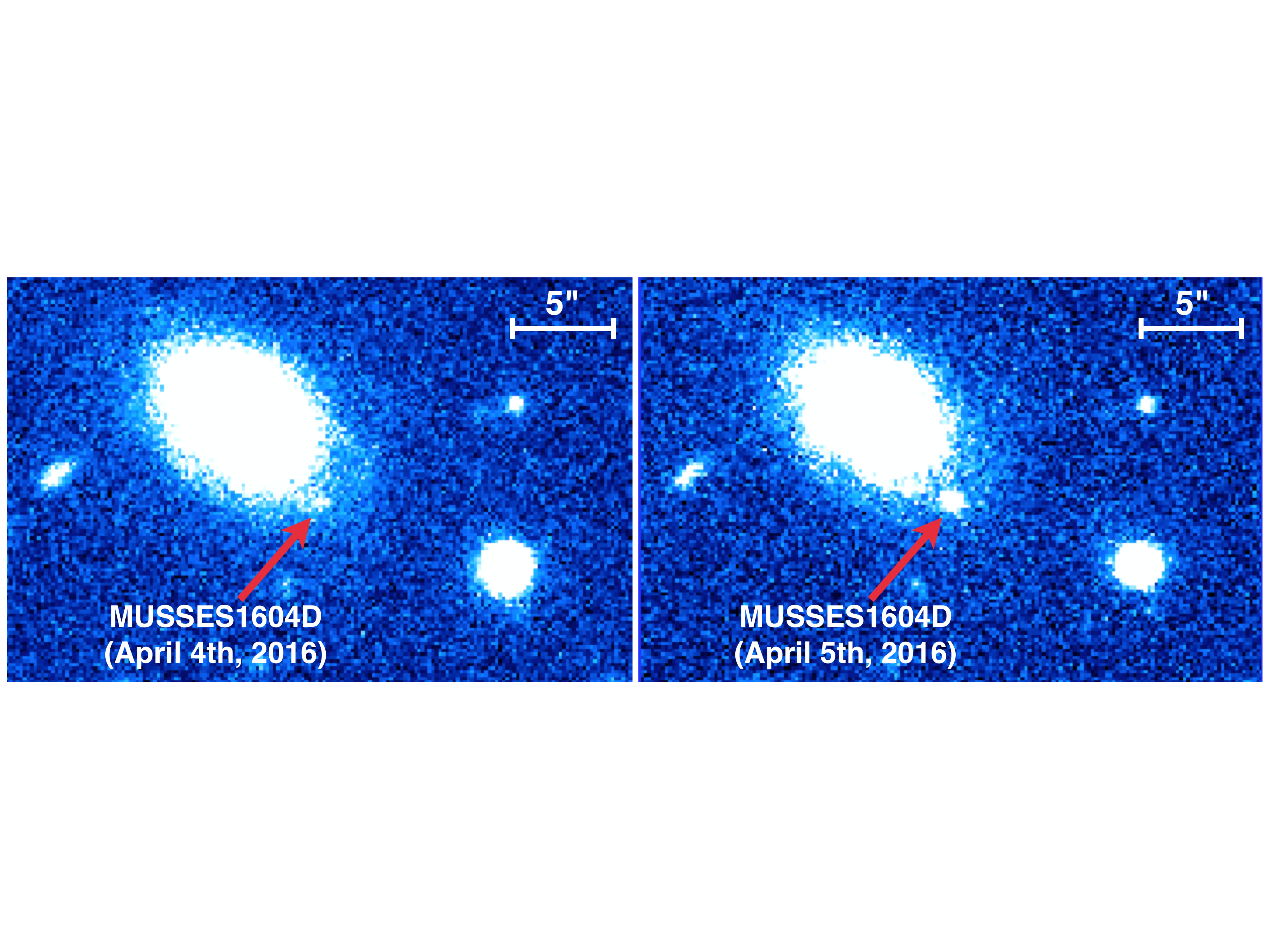}
\centering
\vspace{-80 pt}
\caption*{\textbf{Extended Data Figure 4: Early Subaru/HSC $g$-band images for MUSSES1604D.} The left panel shows the earliest Subaru/HSC image of MUSSES1604D ($\alpha$(J2000) = 12h18m19s.85, $\delta$(J2000) = +00$^\circ$15'17.38") taken on UT April 4.345, 2016, when the $g$-band magnitude of MUSSES1604D was 25.14 $\pm$ 0.15. The supernova then brightened rapidly to $\sim$ 23.1 mag in one day (right panel).}
\end{wrapfigure}

\clearpage

\setlength\intextsep{-11pt}
\begin{wrapfigure}{R}{1\textwidth}
\setlength{\belowcaptionskip}{10pt} 
\includegraphics[width=1\textwidth]{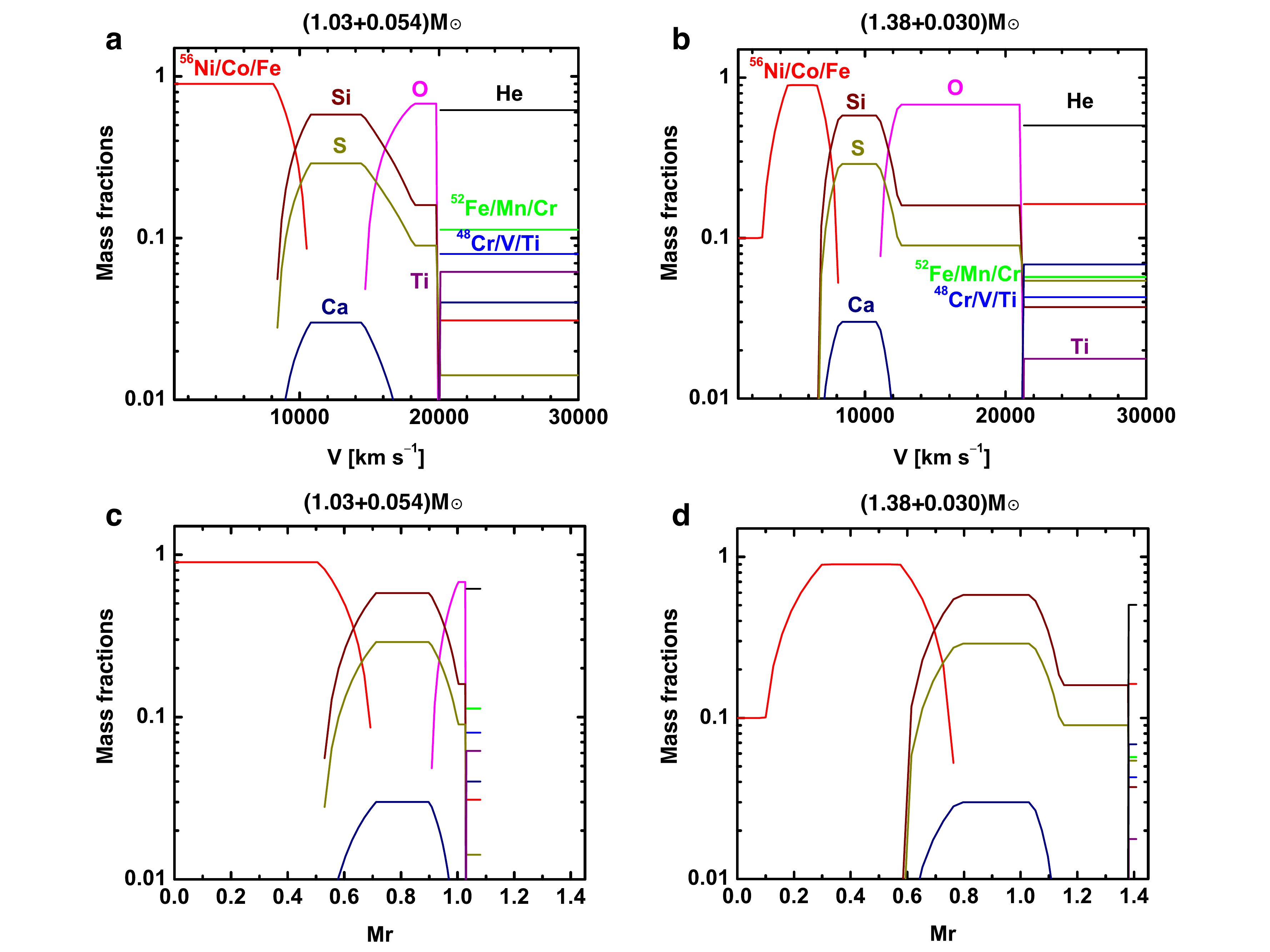}
\centering
\vspace{0 pt}
\caption*{\textbf{Extended Data Figure 5: Composition structures of models used for radiation-transfer simulations.} The composition structures shown here are He-detonation models for the sub-Chandrasekhar-mass WD (1.03 $M_{\odot}$ WD + 0.054 $M_{\odot}$ He-shell; panels \textbf{a} \& \textbf{c}) and the Chandrasekhar-mass WD (1.38 $M_{\odot}$ WD + 0.03 $M_{\odot}$ He-shell; panels \textbf{b} \& \textbf{d}). The mass fractions of selected elements are shown as a function of velocity (\textbf{a}, \textbf{b}) or mass coordinate (\textbf{c}, \textbf{d}). Colors used for selected elements are same for all panels.}
\end{wrapfigure}

\clearpage

\setlength\intextsep{-11pt}
\begin{wrapfigure}{R}{1\textwidth}
\setlength{\belowcaptionskip}{10pt} 
\includegraphics[width=1\textwidth]{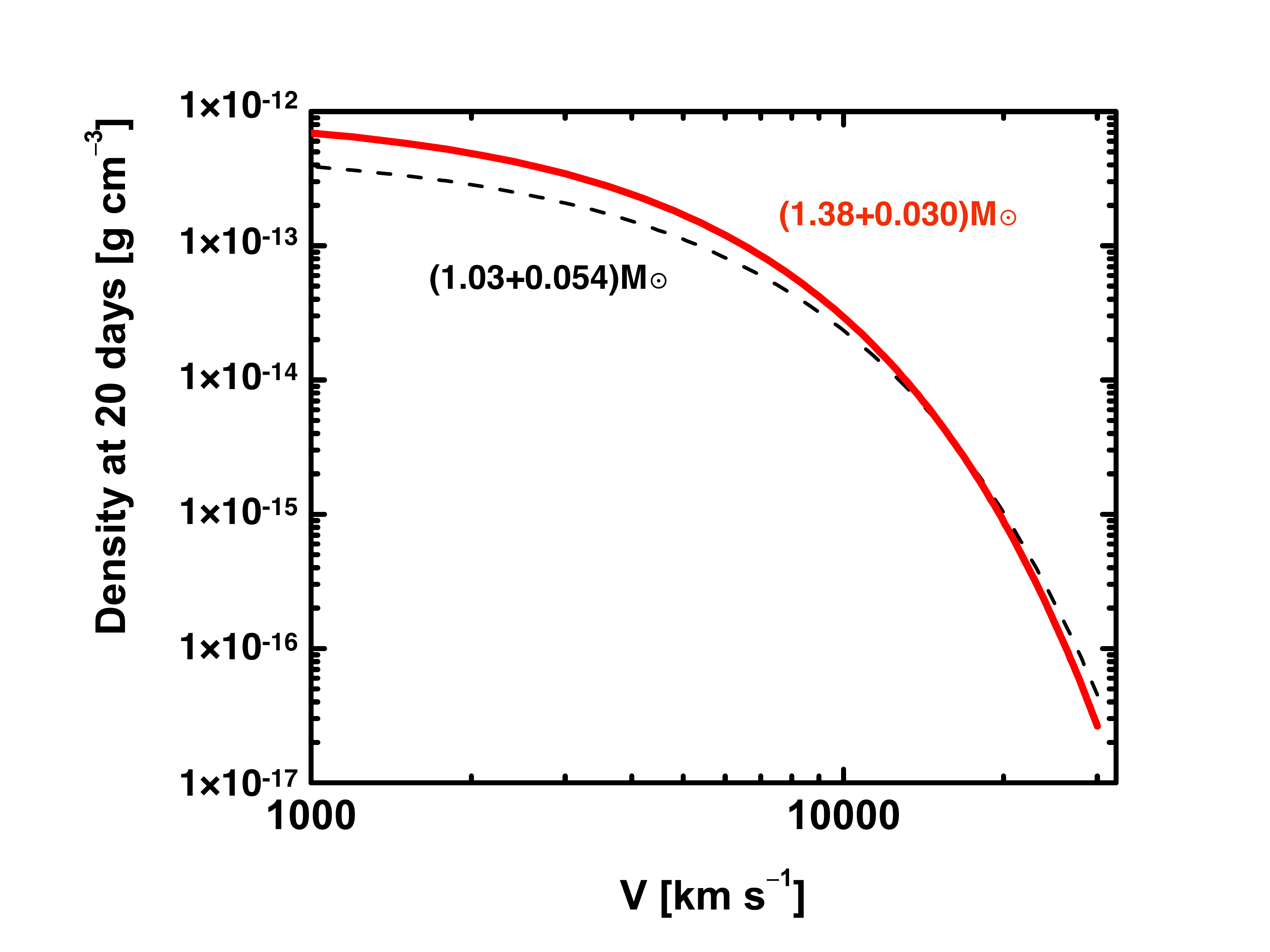}
\centering
\vspace{0 pt}
\caption*{\textbf{Extended Data Figure 6: Density structures of the models used for radiation-transfer simulations.} The density structures (as a function of velocity) shown here are He-detonation models for the sub-Chandrasekhar-mass WD (1.03 $M_{\odot}$ WD + 0.054 $M_{\odot}$ He-shell; black dashed line) and the Chandrasekhar-mass WD (1.38 $M_{\odot}$ WD + 0.03 $M_{\odot}$ He-shell; red solid line).}
\end{wrapfigure}

\clearpage


\begin{table}[htbp] \scriptsize

\sffamily

\begin{center}

\setlength{\belowcaptionskip}{0pt}
\setlength{\abovecaptionskip}{0pt}

\setlength\tabcolsep{3pt}

\begin{threeparttable}[b]

\caption*{\textbf{Extended Data Table 1: Imaging observations of MUSSES1604D}} 

\begin{tabular*}{0.77\paperwidth}{ @{\extracolsep{\fill}}c c c c c c c c c} \hline

\hline

\multirow{2}{*}{UT Date} & \multirow{2}{*}{Phase\tnote{a}} & \multirow{2}{*}{Telescope/Instrument} & \multirow{2}{*}{$g$} & \multirow{2}{*}{$r$} & \multirow{2}{*}{$i$} & \multirow{2}{*}{$B$\tnote{b}} & \multirow{2}{*}{$V$\tnote{b}}\\ 
\\
\hline 

\multirow{2}{*}{Apr 04.34} & \multirow{2}{*}{-19.62} & \multirow{2}{*}{Subaru/HSC} & \multirow{2}{*}{25.14 (15)} & \multirow{2}{*}{--} & \multirow{2}{*}{--} & \multirow{2}{*}{--} & \multirow{2}{*}{--} \\ \\

\multirow{2}{*}{Apr 04.39} & \multirow{2}{*}{-19.58} & \multirow{2}{*}{Subaru/HSC} & \multirow{2}{*}{24.96 (16)} & \multirow{2}{*}{--} & \multirow{2}{*}{--} & \multirow{2}{*}{--} & \multirow{2}{*}{--} \\ \\

\multirow{2}{*}{Apr 05.25} & \multirow{2}{*}{-18.82} & \multirow{2}{*}{Subaru/HSC} & \multirow{2}{*}{23.09 (07)} & \multirow{2}{*}{--} & \multirow{2}{*}{--} & \multirow{2}{*}{--} & \multirow{2}{*}{--} \\ \\

\multirow{2}{*}{Apr 05.29} & \multirow{2}{*}{-18.78} & \multirow{2}{*}{Subaru/HSC} & \multirow{2}{*}{23.15 (06)} & \multirow{2}{*}{--} & \multirow{2}{*}{--} & \multirow{2}{*}{--} & \multirow{2}{*}{--} \\ \\

\multirow{2}{*}{Apr 05.33} & \multirow{2}{*}{-18.74} & \multirow{2}{*}{Subaru/HSC} & \multirow{2}{*}{23.09 (05)} & \multirow{2}{*}{--} & \multirow{2}{*}{--} & \multirow{2}{*}{--} & \multirow{2}{*}{--} \\ \\

\multirow{2}{*}{Apr 05.34 -- 05.43} & \multirow{2}{*}{-18.74 -- -18.65} & \multirow{2}{*}{Subaru/HSC} & \multirow{2}{*}{23.08 (05)} & \multirow{2}{*}{22.99 (05)} & \multirow{2}{*}{--} & \multirow{2}{*}{23.16 (04)} & \multirow{2}{*}{22.97 (05)} \\ \\

\multirow{2}{*}{Apr 05.49} & \multirow{2}{*}{-18.59} & \multirow{2}{*}{Subaru/HSC} & \multirow{2}{*}{--} & \multirow{2}{*}{22.96 (06)} & \multirow{2}{*}{--} & \multirow{2}{*}{--} & \multirow{2}{*}{--} \\ \\

\multirow{2}{*}{Apr 05.55} & \multirow{2}{*}{-18.54} & \multirow{2}{*}{Subaru/HSC} & \multirow{2}{*}{--} & \multirow{2}{*}{22.91 (05)} & \multirow{2}{*}{--} & \multirow{2}{*}{--} & \multirow{2}{*}{--} \\ \\

\multirow{2}{*}{Apr 06.46 -- 06.56} & \multirow{2}{*}{-17.73 -- -17.64} & \multirow{2}{*}{Subaru/HSC} & \multirow{2}{*}{23.05 (05)} & \multirow{2}{*}{22.64 (05)} & \multirow{2}{*}{--} & \multirow{2}{*}{23.08 (04)} & \multirow{2}{*}{22.63 (05)} \\ \\

\multirow{2}{*}{Apr 08.51} & \multirow{2}{*}{-15.89} & \multirow{2}{*}{Subaru/HSC} & \multirow{2}{*}{--} & \multirow{2}{*}{22.14 (04)} & \multirow{2}{*}{--} & \multirow{2}{*}{--} & \multirow{2}{*}{22.50} \\ \\

\multirow{2}{*}{Apr 17.41} & \multirow{2}{*}{-7.93} & \multirow{2}{*}{ARC/ARCTIC} & \multirow{2}{*}{20.35 (03)} & \multirow{2}{*}{20.06 (03)} & \multirow{2}{*}{--} & \multirow{2}{*}{20.33} & \multirow{2}{*}{20.06} \\ \\

\multirow{2}{*}{Apr 26.59} & \multirow{2}{*}{0.29} & \multirow{2}{*}{Kiso/KWFC} & \multirow{2}{*}{--} & \multirow{2}{*}{19.60 (06)} & \multirow{2}{*}{--} & \multirow{2}{*}{19.93} & \multirow{2}{*}{19.61} \\ \\

\multirow{2}{*}{Apr 28.93} & \multirow{2}{*}{2.38} & \multirow{2}{*}{INT/WFC} & \multirow{2}{*}{19.95 (02)} & \multirow{2}{*}{19.64 (01)} & \multirow{2}{*}{19.84 (01)} & \multirow{2}{*}{19.98} & \multirow{2}{*}{19.61} \\ \\

\multirow{2}{*}{May 05.07} & \multirow{2}{*}{7.87} & \multirow{2}{*}{NOT/ALFOSC} & \multirow{2}{*}{20.32 (03)} & \multirow{2}{*}{19.81 (02)} & \multirow{2}{*}{20.04 (02)} & \multirow{2}{*}{20.24} & \multirow{2}{*}{19.76} \\ \\

\multirow{2}{*}{May 05.93} & \multirow{2}{*}{8.64} & \multirow{2}{*}{LT/IO:O} & \multirow{2}{*}{20.39 (02)} & \multirow{2}{*}{19.73 (01)} & \multirow{2}{*}{20.01 (02)} & \multirow{2}{*}{20.29} & \multirow{2}{*}{19.79} \\ \\

\multirow{2}{*}{May 11.93} & \multirow{2}{*}{14.01} & \multirow{2}{*}{LT/IO:O} & \multirow{2}{*}{20.89 (03)} & \multirow{2}{*}{20.04 (03)} & \multirow{2}{*}{20.30 (04)} & \multirow{2}{*}{20.77} & \multirow{2}{*}{20.11} \\ \\

\multirow{2}{*}{May 14.14} & \multirow{2}{*}{15.99} & \multirow{2}{*}{ARC/ARCTIC} & \multirow{2}{*}{21.04 (04)} & \multirow{2}{*}{20.15 (02)} & \multirow{2}{*}{20.29 (02)} & \multirow{2}{*}{20.97} & \multirow{2}{*}{20.22} \\ \\

\multirow{2}{*}{May 19.89 -- 19.92} & \multirow{2}{*}{21.14 -- 21.16} & \multirow{2}{*}{NOT/ALFOSC} & \multirow{2}{*}{21.63 (06)} & \multirow{2}{*}{20.42 (02)} & \multirow{2}{*}{20.32 (04)} & \multirow{2}{*}{21.49} & \multirow{2}{*}{20.43} \\ \\

\multirow{2}{*}{May 25.32} & \multirow{2}{*}{26.00} & \multirow{2}{*}{Gemini-N/GMOS} & \multirow{2}{*}{22.01 (02)} & \multirow{2}{*}{20.67 (02)} & \multirow{2}{*}{--} & \multirow{2}{*}{21.93} & \multirow{2}{*}{20.62} \\ \\

\multirow{2}{*}{May 28.91} & \multirow{2}{*}{29.21} & \multirow{2}{*}{LT/IO:O} & \multirow{2}{*}{22.41 (08)} & \multirow{2}{*}{20.80 (03)} & \multirow{2}{*}{20.55 (03)} & \multirow{2}{*}{22.17} & \multirow{2}{*}{20.78} \\ \\

\multirow{2}{*}{Jun 06.98} & \multirow{2}{*}{37.33} & \multirow{2}{*}{NOT/ALFOSC} & \multirow{2}{*}{22.69 (04)} & \multirow{2}{*}{21.21 (04)} & \multirow{2}{*}{20.98 (05)} & \multirow{2}{*}{22.68} & \multirow{2}{*}{21.26} \\ \\

\hline

\end{tabular*}

\parskip=0.6\baselineskip

\textbf{Notes.} The magnitudes in $g$, $r$ and $i$ bands (observer-frame; AB system) have been transferred to the standard SDSS photometric system by adopting a color term correction based on field stars. Rest-frame $B$- and $V$-band magnitudes are in the Vega system. Numbers in parenthesis correspond to 1-$\sigma$ statistical uncertainties in units of 1/100 mag.

\begin{tablenotes}
\parskip=0.3\baselineskip

\item [a] Days (rest-frame) relative to the estimated date of the $B$-band maximum, UT April 26.27, 2016.

\item [b] K-correction for the flash-phase (April 4--8) observations is carried out by using the power-law spectral energy distribution models derived from the color of the early flash. For post-flash observations, K-correction is performed according to the best-fitting spectral sequence model of MUSSES1604D. The Galactic extinction (E($B-V$)$_{MW}$ = 0.0263 mag) has been corrected.

\end{tablenotes}

\end{threeparttable}

\end{center}

\end{table}

\clearpage


\begin{table}[htbp] \scriptsize

\sffamily

\begin{center}

\setlength{\belowcaptionskip}{0pt}
\setlength{\abovecaptionskip}{0pt}

\setlength\tabcolsep{3pt}

\begin{threeparttable}[b]

\caption*{\textbf{Extended Data Table 2: Properties of MUSSES1604D- and iPTF14atg-like SNe Ia}} 

\begin{tabular*}{0.77\paperwidth}{ @{\extracolsep{\fill}}c c c c c c c c c} \hline

\hline\hline

SN & $B$$_{Max}$\tnote{a} & $\Delta$$m_{15}$($B$) & $B-V$ & Ti$_{II}$ & Si$_{II}$ & $V$$_{Si_{II}\lambda6355}$\tnote{c} & Ti$_{II}$\\ 
Name & (mag) & (mag) & Level\tnote{b} & Absorptions\tnote{c,d} & $\lambda$5972\tnote{c,e} & (km s$^{-1}$) & Evolution\tnote{f}\\ [1ex]

\hline 

\multicolumn{8}{c}{\textbf{MUSSES1604D-Like SN Ia Candidates}} \\

\hline

MUSSES1604D & -18.8$_{+}$ & 1.0$_{+}$ & Red$_{+}$ & Very Deep$_{+}$ & Shallow$_{+}$ & 11,800$_{+}$  & Slow$_{+}$ \\

SN~2012df$_{+}$ & -18.9\tnote{g}$_{+}$ &---~~ & Marginal-red$_{+}$ & Deep$_{+}$ & Shallow$_{+}$ & 12,000$_{+}$ & ---~~ \\

SN~2007cq$_{+}$ & -19.4$_{+}$ & 1.1$_{+}$ & Normal$_{-}$ & Deep$_{+}$ & Shallow$_{+}$ & 11,000$_{+}$ & Slow$_{+}$ \\

SN~2006bt$_{+}$ & -18.9$_{+}$ & 1.1$_{+}$ & Marginal-red$_{+}$ & Deep$_{+}$ & Intermediate $_{-}$ & 11,600$_{+}$ & Slow$_{+}$ \\

\hline

SN~2011fe$_{-}$ & -19.2$_{+}$ & 1.2$_{~|}$ & Normal$_{-}$ & Intermediate$_{-}$ & Intermediate $_{-}$ & 10,300$_{~|}$ & Normal$_{-}$ \\

\hline

\multicolumn{8}{c}{\textbf{iPTF14atg-Like SN Ia Candidates}} \\

\hline

iPTF14atg & -17.7$_{+}$ & 1.3$_{+}$ & Red$_{+}$ & Very Deep$_{+}$ & Deep$_{+}$ & 7,300$_{+}$ & Normal$_{+}$ \\

SN~2002es$_{+}$ & -17.9$_{+}$ & 1.3$_{+}$ & Red$_{+}$ & Very Deep$_{+}$ & Deep$_{+}$ & 6,000$_{+}$ & Normal$_{+}$ \\

\hline

SN~2010lp$_{?}$ & -17.9$_{+}$ & 1.4$_{+}$ & Ultra-Red$_{+}$ & Very Deep$_{+}$ & Deep$_{+}$ & 10,600$_{-}$ & ---~~ \\

PTF10ops$_{?}$ & -17.8$_{+}$ & 1.1$_{+}$ & Ultra-Red$_{+}$ & Deep$_{+}$ & Deep$_{+}$ & 10,000$_{-}$ & Normal$_{+}$ \\

SN~2011ay$_{?}$ & -18.1$_{+}$ & 1.3$_{+}$ & Red$_{+}$ & Deep$_{+}$ & Intermediate$_{~|}$ & 5,600$_{+}$ & Fast$_{-}$ \\

SN~2008A$_{?}$ & -17.9$_{+}$ & 1.3$_{+}$ & Red$_{+}$ & Deep$_{+}$ & Intermediate $_{~|}$ & 6,900$_{+}$ & Fast$_{-}$ \\

SN~2005hk$_{?}$ & -17.7$_{+}$ & 1.5$_{~|}$ & Marginal-red$_{~|}$ & Deep$_{+}$ & Deep $_{+}$ & 6,100$_{+}$ & Fast$_{-}$ \\

SN~2008ae$_{?}$ & -17.1$_{~|}$ & 1.4$_{+}$ & Ultra-Red$_{+}$ & Deep$_{+}$ & Intermediate $_{~|}$ & 7,900$_{+}$ & Fast$_{-}$ \\

SN~2002cx$_{?}$ & -17.5$_{+}$ & 1.2$_{+}$ & Normal$_{-}$ & Deep$_{+}$ & Intermediate $_{~|}$ & 5,500$_{+}$ & Normal$_{+}$ \\

\hline\hline

\end{tabular*}

\parskip=1.0\baselineskip

\textbf{Notes.} For each property, we use ``+", ``$~|~$" and ``-" footnotes as ``support", ``neutral" and ``opposite" respectively to show the similarity between candidates and MUSSES1604D/iPTF14atg. For all three MUSSES1604D-like SNe Ia, the host extinction are neglected because of the relatively distant location of SNe to  the center of their S0/a host galaxies and the non-detection of Na~I~D lines in their spectra. Galactic extinction has been applied with E($B-V$)$_{MW}$ of 0.1096 mag and 0.050 mag for SN~2007cq and SN~2006bt respectively.

\begin{tablenotes}
\parskip=0.3\baselineskip

\item [a] The absolute magnitude for iPTF14atg, 02es-like (SN~2002es, SN~2010lp, PTF10ops) and all normal-brightness SNe Ia was calculated by using cosmological parameters H$_{0}$ = 70.0 km s$^{-1}$ Mpc$^{-1}$, $\Omega$$_{m}$ = 0.30, $\Omega$$_{\Lambda}$ = 0.70 and $\Omega$$_{\nu}$ = 0.00. For 02cx-like SNe Ia, we adopt the value from the related paper$^{68}$.
\item [b] The $B-V$ color information at around the $B$-band maximum. Here, we define $B-V$ $\ge$ 0.4 mag, 0.4 mag $>$ $B-V$ $\ge$ 0.2 mag, 0.2 mag $>$ $B-V$ $\ge$ 0.1 mag, 0.1 mag $>$ $B-V$ $\ge$ -0.1 mag and -0.1 mag $>$ $B-V$ as ``ultra-red", ``red", ``marginal-red", ``normal" and ``blue", respectively. 
\item [c] Spectral features at around the $B$-band maximum. For normal-brightness and subluminous SNe Ia, we used spectra taken on the closest epoch to t = -2 and t = 0, respectively (relative to the $B$-band maximum) for the similarity comparisons. 
\item [d] The relative strength of Ti II absorptions near the $B$-band maximum. The strength is relative to normal-type SNe Ia, e.g. SN~2011fe. 
\item [e] We define the equivalent width (EW) of Si II $\lambda$5972 line as: EW (Si${_{II\lambda5972}}$) $\le$ 10 $\AA$, 10 $\AA$ $<$ EW (Si${_{II\lambda5972}}$) $\le$ 30 $\AA$, 30 $\AA$ $<$ EW (Si${_{II\lambda5972}}$) as ``Shallow", ``Intermediate" and ``Deep", respectively.
\item [f] The relative evolution speed of Ti II absorptions in the first 10 $\pm$ 2 days after the $B$-band maximum. The evolution speed is relative to SN~2011fe and iPTF14atg for normal-brightness and subluminous SNe Ia respectively.
\item [g] Unfiltered photometry without considering the Galactic extinction E($B-V$)$_{MW}$ = 0.0393 mag.

   \end{tablenotes}
\end{threeparttable}

\label{table:Properties of MUSSES1604D- and iPTF14atg-like SNe Ia} 

\end{center}

\end{table} 
\clearpage


\end{document}